\documentclass[%
 reprint,
 amsmath,amssymb,
pra,
]{revtex4-2}
\usepackage{graphicx}
\usepackage{caption}
\usepackage{subfigure}
\usepackage{color}
\usepackage{graphicx}
\usepackage{dcolumn}
\usepackage{bm}

\usepackage{amsfonts,amsmath,amssymb,mathrsfs,amsthm}
\usepackage{epsfig,epstopdf,graphicx,graphics}
\usepackage{color}
\usepackage{float}
\usepackage{ifpdf}
\usepackage[colorlinks=true]{hyperref}
\usepackage{eucal}
\usepackage{braket}
\usepackage{graphicx}
\usepackage{dcolumn}
\usepackage{bm}
\begin{document}
\preprint{APS/123-QED}
\title{Amplification of optical Schr\"{o}dinger cat states with implementation protocol based on frequency comb}

\author{Hongbin Song}
 \email{songhongbin@cuhk.edu.cn}
 \affiliation{General Education Division, The Chinese University of Hong Kong, Shenzhen, \\Guang Dong 518116, China\\ }%

\author{Guofeng Zhang}
\email{guofeng.zhang@polyu.edu.hk}
\author{Xiaoqiang Wang}
\affiliation{Department of Applied Mathematics, The Hong Kong Polytechnic University, Kowloon 999077, Hong Kong, China}%
\affiliation{
 The Hong Kong Polytechnic University Shenzhen Research Institute, Shenzhen 518057, China\\
}%
\author{Hidehiro Yonezawa}
\email{h.yonezawa@unsw.edu.au}
\affiliation{%
 Centre for Quantum Computation and Communication Technology and\\ School of Engineering and Information Technology,The University of New South Wales,\\
  Canberra, ACT 2600, Australia\\}
 \author{Kaiquan Fan}
  \affiliation{School of Science and Engineering, The Chinese University of Hong Kong, Shenzhen 518116, China}%
\date{\today}

\begin{abstract}
We proposed and analyzed a scheme to generate large-size Schr\"{o}dinger cat states based on linear operations of Fock states and squeezed vacuum states and conditional measurements. By conducting conditional measurements via photon number detectors, two unbalanced Schr\"{o}dinger kitten states combined by a beam splitter can be amplified to a large-size cat state with the same parity. According to simulation results,  two Schr\"{o}dinger odd kitten states of $\beta=1.06$ and $\beta=1.11$ generated from one-photon-subtracted squeezed vacuum states of $-$3 dB, are amplified to an odd cat state of $\beta=1.73$ with a fidelity of $F=99\%$. A large-size Schr\"{o}dinger odd cat state with $\beta=2.51$  and $F=97.30\%$ is predicted when the input squeezed vacuum states are increased to $-$5.91 dB. According to the analysis on the impacts of experimental imperfections in practice, Schr\"{o}dinger odd cat states of $\beta>2$ are available. A feasible configuration based on a quantum frequency comb is developed to realize the large-size cat state generation scheme we proposed. 
\end{abstract}

\maketitle


\section{Introductaion}

Optical Schr\"{o}dinger cat states described as $|\beta\rangle+e^{i\phi}|-\beta\rangle$ for even cats and odd cats corresponding to $\phi=0$ and $\phi=\pi$ respectively, are the superposition of two distinguishable coherent states $|\beta\rangle$ and $|-\beta\rangle$ with opposite phases, where $|\beta|$ describes the size of Schr\"{o}dinger cat states  \cite{PRADakna1997}. Odd(Even) cats are featured with odd(even) photon number distribution. Different from the mixture states, superposition states reveal the interference between superposed components, which play an important role in the verification of quantum nonlocality \cite{PRAJeong2003},  quantum communication \cite{JoptBSangouard2010,PRLBrask2010,NPhotonicsJonas2013}, continuous-variable quantum computation \cite{PRACochrane1999,OLHastrup2020, PRARalph2003, PRLLund2007, JOptBGilchristy2004} and quantum metrology \cite{PRAMunro2002,PRLMunro2011}. 
       One-photon subtracted squeezed vacuum state is a standard method to generate  Schr\"{o}dinger cat states with small size, which is defined as Schr\"{o}dinger kitten states  \cite{PRADakna1997}. Kitten states at baseband  \cite{ScienceOurjoumtsev2006, PRLNielsen2006, OptExpressWakui2007, NatPhotonicsNamekata2010}  and sidebands\cite{PRLSerikawa2018} have been generated. While the maximum amplitude of cat states generated with such method is limited as $|\beta|=1.2$ to keep the fidelity as high as $99\%$ \cite{PRLLund2004,PRAAgata2008}.  Fidelity between an ideal cat state $|\phi\rangle$ and a pure state $|\psi\rangle$ is usually defined as F$(\phi,\psi)=|\langle\phi|\psi\rangle|^2$ \cite{NatureOurjoumtsevl2007, PRLLund2004} (the definition in the form of square root, i.e. $\langle\phi|\psi\rangle$ is also used  in some references such as \cite{NatphotonSychev2017,PRAAgata2008}), indicating the similarity of the generated state with an ideal cat state, which is an important quality measure of the cat state.  
      
    However, the overlap between two superposed coherent states in a Schr\"{o}dinger cat state, i.e. $\langle\beta|-\beta\rangle=e^{-2|\beta|^{2}}$, is required to approach zero to effectively work as qubits in quantum information science, i.e. $|\beta|\geq2$ \cite{PRARalph2003}. Therefore, large-size Schr\"{o}dinger cat states generation has attracted intense interest.
    
    To create Schr\"{o}dinger cat states with larger amplitude, theoretical and experimental investigations on multi-photon subtracted squeezed vacuum states were conducted \cite{PRLTakahashi2008, PRAMasahirohiroki2008, PRANielsen2007, PRAGerrits2010}. An even cat state and an odd cat state of $\beta=1.4$ and $F=60\%$ as well as $\beta=1.7$ and $F=59\%$ were experimentally created by subtracting two and three photons from squeeze vacuum states \cite {PRLTakahashi2008,PRAGerrits2010}, respectively. Based on a Fock state and homodyne detection, a squeezed even cat state \cite{PRAMasahirohiroki2008} with $|\beta|=\sqrt{2.6}\approx 1.61$ was experimentally demonstrated \cite{NatureOurjoumtsevl2007}. An iterative scheme to amplify Schr\"{o}dinger kitten states was proposed by conducting auxiliary-coherent-state-aided conditional measurement on two combined kitten states by a 50/50 beam splitter \cite{PRLLund2004}, from which a high-fidelity ($F\geq99\%$) cat state with $\beta=2.5$ was predicted with four iterative stages and inefficient photon detection. This approach was extended to a homodyne heralding scheme later, in which the conditional measurement with photon detectors and auxiliary coherent state was replaced with homodyne detection \cite{PRALaghaout2013}. Two balanced Schr\"{o}dinger odd kitten states with $|\beta|=1.15$ squeezed by 1.74 dB were amplified to a Schr\"{o}dinger even cat state of $|\beta|=1.85$ squeezed by 3.04 dB with a fidelity of $F= 59.29\%$ (in the definition of $F=|\langle \phi|\psi\rangle|^2$) based on the homodyne heralding scheme \cite{NatphotonSychev2017}, which was considered as the best experimental result till now \cite{PRATakase2021}. A sequential photon catalysis scheme was proposed to generate large-size squeezed Schr\"{o}dinger cat states \cite{NJPLEaton2019}, in which multi-stage photon catalyses are required to breed the cat state. Large-size Schr\"{o}dinger cat states of $|\beta|\textgreater2$ with a fidelity around $F= 99\%$ based on a two-mode 9-photon entangled state were predicted \cite{ScientificreportMilheev2019}. Recently, an optical cat state generation scheme based on the general photon subtraction of two squeezed vacua was reported with high generation rate, in which a large-size squeezed cat state of $|\beta|=\sqrt{10}\approx3.16$ was predicted with a fidelity of $F=99.7\%$ by subtracting 10 photons from one mode of the two-mode squeezed vacuum states \cite {PRATakase2021}. However, the operation of multi-photon subtraction or iterative/sequential process significantly increases the complexity of the experimental setup in practice.
   \newline 
   \newline 
     In this paper, we propose a scheme to generate large-size odd cat states based on linear operations of squeezed vacuum states and conditional measurements. Amplification of odd kitten states that keeps the same parity is realized. Schr\"{o}dinger odd cat states of $\beta \textgreater 2$ is predicted with one-photon subtraction through a single run. We also develop a quantum-frequency-comb-based protocol to realize the scheme. \\
        \newline 
     The paper is organized as follows. In section \ref{general-l-k-sqzvac}, a general model for $l$-added-and-$k$-subtracted squeezed vacuum states is introduced based on tensor operation. In section \ref{amplificationmodel}, an effective approach to producing large-size Schr\"{o}dinger cat states is developed. The impacts of imperfect kitten states and photon number detectors are analyzed in section \ref{model with imperfection}. In section \ref{frequency comb-based protocol}, a protocol for the experimental implementation of the scheme based on a quantum frequency comb is proposed. Concluding remarks are provided in section \ref{conclusion}. 
    
\section{General model for cat state production with $l$-added-and-$k$-subtracted squeezed vacuum states}\label{general-l-k-sqzvac} 
\subsection{Schr\"{o}dinger cat states}
 Optical Schr\"{o}dinger cat states, i.e. the superposition of two coherent states with opposite phases, $|\beta\rangle$ and $|$-$\beta$$\rangle$, can be written as \cite{PRADakna1997}
  \begin{equation}
  |\psi_{\mathrm{cat}}\rangle=N_{\pm}(|\beta\rangle\pm|-\beta\rangle),
                                                                       \label{equ1cat}                                                 
\end{equation}
where, 
  \begin{equation}
  N_{\pm}=\frac{1}{\sqrt{2(1\pm e^{-2|\beta|^{2}})}},
  \label{equ2_normaliztaion}
\end{equation}
    
    \begin{equation}
  |\beta\rangle=e^\frac{-|\beta|^2}{2}\sum_{n=0}^{\infty}\frac{\beta^n}{\sqrt{n!}}|n\rangle,
                                                         \label{equ1}
\end{equation}
in which $|n\rangle$ is a Fock state. ``$+$" and ``$-$" represent even and odd Schr\"{o}dinger cat states, respectively. $N_{\pm}$ are the corresponding normalization constants. Substituting Eq. (\ref{equ1}) to Eq. (\ref{equ1cat}), it can be seen that only even(odd) photons are distributed in even(odd) cat states. 
 To ensure two superposed coherent states are orthogonal, the overlap between $|\beta\rangle$ and $|-\beta\rangle$ should approach zero \cite{PRARalph2003}, i.e. 
\begin{equation}
      \langle\beta|-\beta\rangle=e^{-|\beta|^2}\sum_{n=0}^{\infty}\frac{(-\beta^2)^n}{n!}\\
                                                    =e^{-2|\beta|^2} \ll0,     
                                                                                 \label{equ4}
\end{equation}
when $|\beta|\geq2$. 
 \subsection{General model of \textit{l}-added-and-\textit{k}-subtracted squeezed vacuum states}
  \begin{figure}[htbp]
  \centering
  \includegraphics[width=8cm]{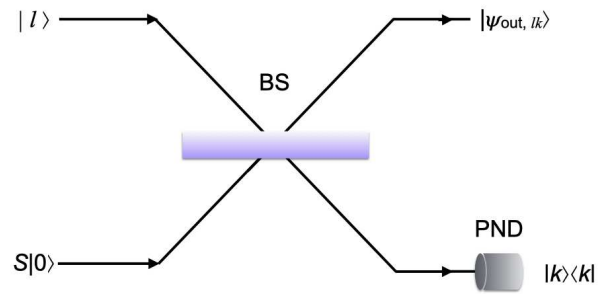}
  \caption{Schematic diagram of Schr\"{o}dinger cat state \\
  generation based on $l$-added-and-$k$-subtracted \\
  squeezed vacuum states\\
 BS: Beam splitter;  PND: Photon number detector}
  \label{l-added-k-sub}
\end{figure}
Similar to an even cat state, a squeezed vacuum state only contains even-photon number distribution, which  is written as  \cite{PRADakna1997,PRAAgata2008},
    \begin{equation}
\hat{S}(\xi)|0\rangle=\sum_{n=0}^{\infty}\alpha _{2n}|2n\rangle ,
\label{eq:aug10_squeezed_vacuum}
\end{equation}%
where%
\begin{equation}
\alpha _{2n}=\frac{1}{\sqrt{\cosh \xi }}\frac{\sqrt{(2n)!}\tanh
^{n}\xi }{2^{n}n!}, \label{eq:aug10_alpha_2n}
\end{equation}%
and $\xi$ is the squeezing parameter. 
Thus, squeezed vacuum states is a high-fidelity  ($F\geq99\%$) approximation to the even kitten states ($\beta\leq0.75$) \cite{PRAAgata2008}. Odd-photon number distribution can be obtained if odd photons such as 1 or 3 are added/subtracted from a squeezed vacuum state. So it is an effective approach to generating Schr\"{o}dinger kitten states. Here, we generalize the model to \textit{l}-added-and-\textit{k}-subtracted squeezed vacuum states as shown in Fig. \ref{l-added-k-sub}. Such model is a quantum linear system driven by multi-channel multi-photon states \cite{IEEETRANONAUTOCONTROLZhang2013, Automaticzhang2014,Automaticzhang2017}. The total input state of the system can be written as, 
\begin{eqnarray}
|\Psi _{\mathrm{in}}\rangle &=&|l\rangle\otimes\sum_{n=0}^{\infty }\alpha _{2n} |2n\rangle \\ \nonumber
\label{Totalinput}
\end{eqnarray}
 According to tensor operation of quantum linear system driven by multi-channel multi-photon states \cite{IEEETRANONAUTOCONTROLZhang2013, Automaticzhang2014,Automaticzhang2017}, the generated state at the output of the beam splitter can be derived as,
 \begin{widetext}
\begin{eqnarray}
|\Psi_{\mathrm{out}}\rangle&=&\sum_{n=0}^{\infty }\frac{\alpha _{2n}}{\sqrt{\ell!(2n)!}}\sum_{i=0}^{\ell}\sum_{j=0}^{2n}\binom{\ell }{i}\binom{2n}{2n-j} \sqrt{(\ell +j-i)!(2n+i-j)!}
\nonumber\\
&&(-1)^{j}T^{2n+\ell -i-j}R^{i+j}|\ell +j-i\rangle \otimes |2n+i-j\rangle ,
\end{eqnarray}
\end{widetext}
where, $R^2$ and $T^2$ are the reflectivity and transmittance of the beam splitter.
When conditional measurement is conducted with a Fock state $|k\rangle$, we have
 \begin{eqnarray}
|\Psi_{\mathrm{out},\;lk}\rangle&=&\frac{1}{N_{n, l k}}\sum_{n=0}^{\infty} \gamma_{n,\;lk} |{2n+l-k}\rangle\nonumber\\
 \label{eq:jun1_psiout_nlk}
\end{eqnarray}

where, 
\begin{eqnarray}
 \gamma_{n,\;l k}
&=&\frac{\alpha _{2n}}{\sqrt{l!(2n)!}}\sum_{j=\max(k-l,0)}^{\min(k, 2n)}\binom{l }{ l-k +j}\binom{2n}{2n-j}
\nonumber\\
&&\sqrt{k!(2n+l-k)!}%
(-1)^{j}T^{2n+k-2j}R^{l-k+2j},
 \label{eq:jun1_gamma_nlk}
\end{eqnarray}
and $N_{n,\; l k}$ is a normalization constant. The density matrix of the heralded state in Eq. (\ref{eq:jun1_psiout_nlk}) is described as $\rho_{out,\;l k}=|\Psi_{out,\; l k}\rangle\langle\Psi_{out,\; l k}|$.
 The fidelity between two mixed states with density matrices $\rho_{1}$ and $\rho_{2}$ is generally defined as 
\begin{equation}
F(\rho_1,\rho_2)=(\mathrm{Tr}[(\sqrt{\rho_{1}}\rho_{2}\sqrt{\rho_{2}})^\frac{1}{2}])^2,
\label{eq:generalfidelity}
\end{equation}
where, $\mathrm{Tr}$[\; ] denotes the trace of a matrix \cite{JOMJozsa1994}. In the case of Schr\"{o}dinger cat state generation, an ideal pure cat state $|\psi_{cat}\rangle$ is always used to calculate the fidelity. Thus. Eq. (\ref{eq:generalfidelity}) can be rewritten as, 
\begin{equation}
F(\rho_{\mathrm{out},\;l k},\psi_{\mathrm{cat}}) \triangleq\langle \psi_{\mathrm{cat}}|\rho_{\mathrm{out},\;l k}|\psi_{\mathrm{cat}}\rangle.
\end{equation}

\begin{figure*} [htbp]
\centering
\subfigure[]{
\label{fig:subfig:a xi0.68beta1.991} 
\includegraphics[height=2.4in,width=2.95in]{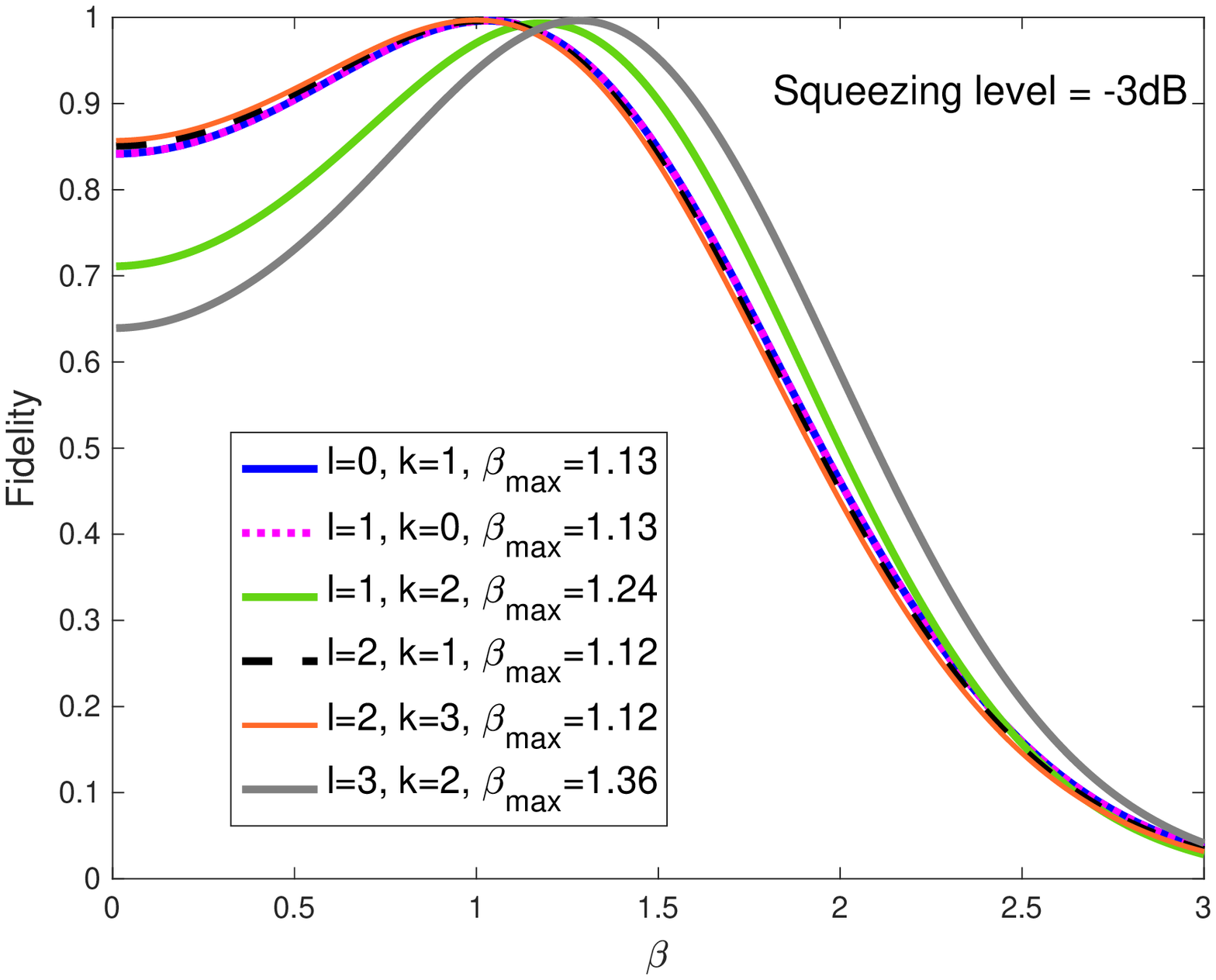}}
\subfigure[]{ \label{fig:subfig:b xi0.7beta2} 
\includegraphics[height=2.4in,width=2.95in]{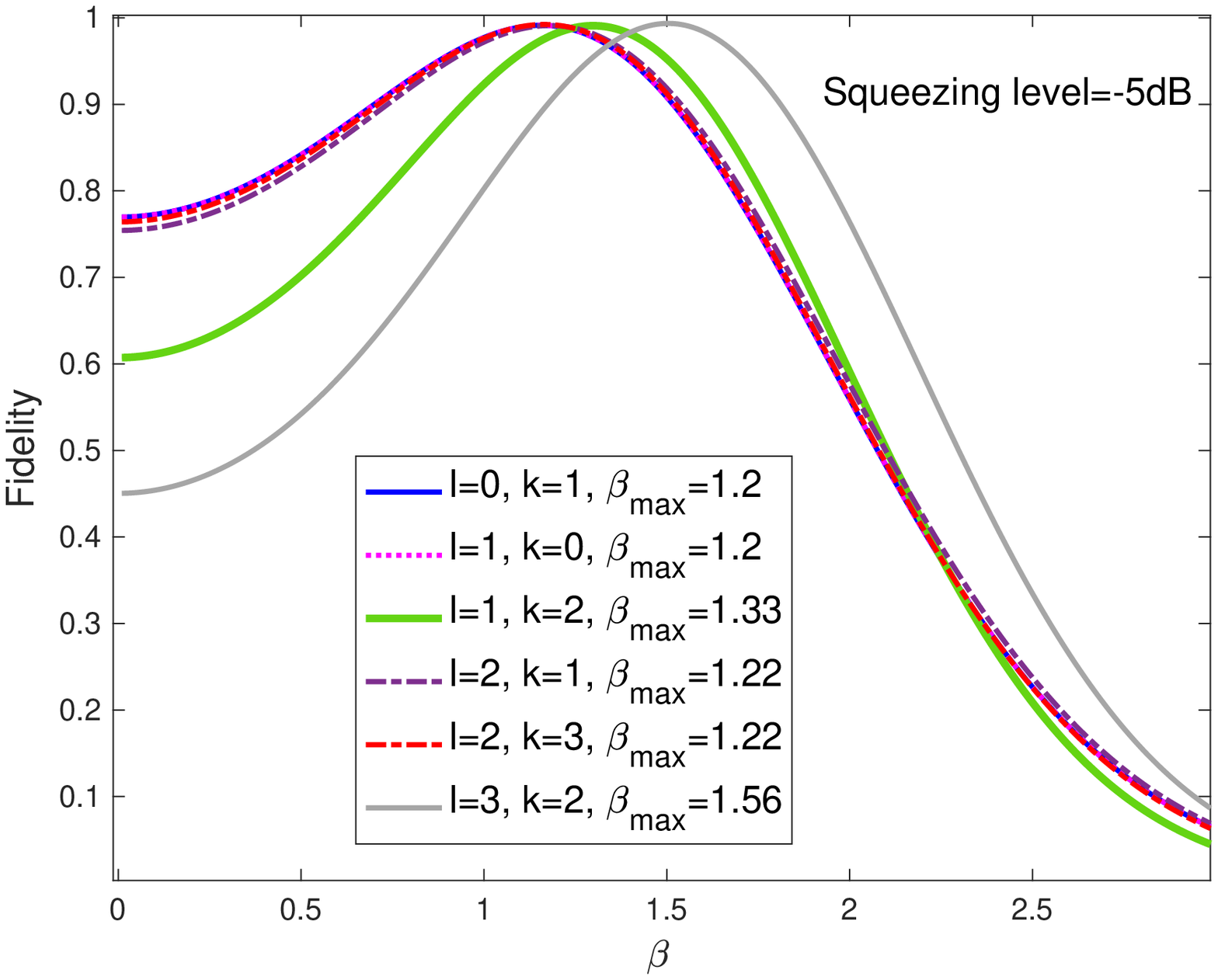}}
 \caption{Fidelity between an ideal odd cat state and the generated cat states\\
 from $l$-added and $k$-subtracted squeezed vacuum states\\
(a) squeezing level of -3 dB (b) squeezing level of -5 dB}
\label{comparison_F_beta_for_diff_L_l} 
\end{figure*}
When $\textit{l}-\textit{k}$ is odd, an odd cat state is produced, while an even cat state is generated when $l-k$ is even. The case, $l=0$ and $k=1$, i.e. one-photon-subtracted squeezed vacuum state, has become the standard approach to generating odd kitten states \cite{ScienceOurjoumtsev2006, PRLNielsen2006, OptExpressWakui2007, NatPhotonicsNamekata2010}. Fig. \ref{comparison_F_beta_for_diff_L_l} depicts the fidelity between an ideal Schr\"{o}dinger odd cat state and the generated odd cat states from $l$-added-and-$k$-subtracted squeezed vacuum states with the squeezing levels of -3 dB and -5 dB. The reflectivity of the beam splitter is numerically optimized by the fidelity. It is verified that the size of high-fidelity ($F\geq99\%$) kitten states generated from one-photon subtracted squeezed vacuum state is limited as $\beta$=1.2 \cite{PRLLund2004,PRAAgata2008}, even though the squeezed vacuum state is increased to -5 dB. In addition, the performance of one-photon subtracted squeezed vacuum state is verified to be similar to that of one-photon added squeezed vacuum state. All schemes show that the fidelity increases with $\beta$ before it reaches the maximum and then it drops with the further increase of $\beta$. Here, $\beta_{max}$ is taken as the maximum value of $\beta$ when $F=99\%$. 
Moreover, high-fidelity kitten states with $\beta\textgreater1.2$ can be obtained when $l=1$ and $k=2$ as well as $l=3$ and $k=2$ in the scheme, which break the limit, $\beta=1.2$ \cite{PRLLund2004,PRAAgata2008}, of the standard approach with $l=0$ and $k=1$.  
 \begin{figure}[htbp]
\centering
  \includegraphics[width=8cm]{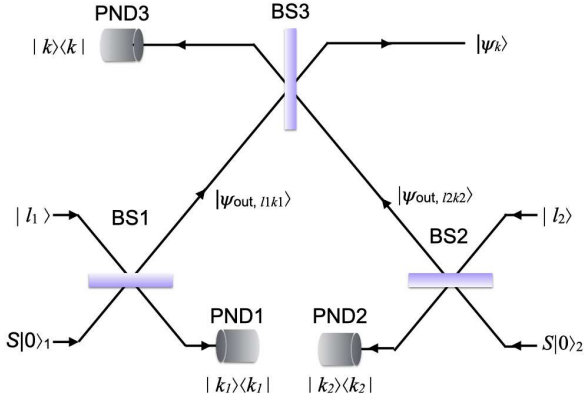}
  \centering \caption{Schematic diagram for large-size Schr\"{o}dinger\\
   cat state generation\\
  BS: Beam splitter; $ \;$  PND: Photon number detector}
  \label{l-added-k-sub-two-BS}
\end{figure}
\section{Model for large-size schr\"{o}dinger cat state generation} \label{amplificationmodel}

An approach to generating large-size Schr\"{o}dinger cat states based on $\textit{l}$-photon-added-and-$\textit{k}$-photon-subtracted squeezed vacuum states is proposed as shown in Fig. \ref{l-added-k-sub-two-BS}. Two in-phase Schr\"{o}dinger kitten states generated with $\textit{l}$-added-and-$\textit{k}$-subtracted squeezed vacuum states are combined on the third beam splitter, BS3. The total input state to BS3 is described as,
\begin{eqnarray}
&&|\Psi^1 _{\mathrm{out},\;l_1k_1}\rangle\otimes|\Psi^2 _{\mathrm{out},\;l_2k_2}\rangle\nonumber\\
&=&\sum_{n=0}^{\infty }\sum_{m=0}^{\infty } \gamma^1_{n,\;\ell_1 k_1}\gamma^2_{m,\;\ell_2 k_2}\nonumber\\
&&|2n+\ell_1-k_1\rangle\otimes|2m+\ell_2-k_2\rangle,  
  \label{eq_gamma12}
\end{eqnarray}

where, $\gamma^1_{n, \;\ell_1 k_1}$ and $\gamma^2_{m,\; \ell_2 k_2}$ have the same formats to $\gamma_{n,\;lk}$ in Eq. (\ref{eq:jun1_gamma_nlk}).
The total input state is a linear combination of products  $|2n+\ell_1-k_1\rangle\otimes |2m+\ell_2 -k_2\rangle$, where $n,m=0,\ldots, \infty$. According to the general theory in \cite{IEEETRANONAUTOCONTROLZhang2013, Automaticzhang2014,Automaticzhang2017}, BS3 implements the following mapping,

\begin{widetext}
\begin{eqnarray}
&&|2n+\ell_1-k_1\rangle\otimes|2m+\ell_2 -k_2\rangle
\nonumber\\
&\rightarrow&
\nonumber
\\
&&\frac{1}{\sqrt{(2n+\ell_1-k_1)! (2m+\ell_2 -k_2)!}}  \sum_{j=0}^{2n+\ell_1-k_1}\sum_{l=0}^{2m+\ell_2 -k_2} \binom{2n+\ell_1-k_1}{j}\binom{2m+\ell_2 -k_2}{l} (-1)^l 
\nonumber
\\
&&T_3^{2m+\ell_2 -k_2+j-l} R_3^{2n+\ell_1-k_1-j+l}\sqrt{(j+l)! (2n+\ell_1-k_1+2m+\ell_2 -k_2-j-l)!}
\nonumber
\\
&&|j+l\rangle|2n+\ell_1-k_1+2m+\ell_2 -k_2-j-l\rangle.
\nonumber
\\
&=&  \sum_{j=0}^{2n+\ell_1-k_1}\sum_{l=0}^{2m+\ell_2 -k_2}\gamma_{n,m,j,l}|j+l\rangle|2n+\ell_1-k_1+2m+\ell_2 -k_2-j-l\rangle,
\end{eqnarray}
where 
\begin{eqnarray}
\gamma_{n,m,j,l} &\triangleq& \frac{(-1)^lT_3^{2m+\ell_2 -k_2+j-l} R_3^{2n+\ell_1-k_1-j+l}}{\sqrt{(2n+\ell_1-k_1)! (2m+\ell_2 -k_2)!}}\binom{2n+\ell_1-k_1}{j}\binom{2m+\ell_2 -k_2}{l} \nonumber\nonumber\\
 &&\sqrt{(j+l)! (2n+\ell_1-k_1+2m+\ell_2 -k_2-j-l)!}.
   \label{eq:jun1_gamma_nmjk}
\end{eqnarray}
\end{widetext}
Consequently, the final output state of BS3 is

\begin{eqnarray}
|\Psi\rangle &=& \sum_{n=0}^{\infty }\sum_{m=0}^{\infty} \gamma^1_{n,\;\ell_1 k_1}\gamma^2_{m, \;\ell_2 k_2}  \sum_{j=0}^{2n+\ell_1-k_1}\sum_{l=0}^{2m+\ell_2 -k_2}\gamma_{n,m,j,l}\nonumber\\
&&|j+l\rangle|2n+\ell_1-k_1+2m+\ell_2 -k_2-j-l\rangle.
\label{eq:jun1_outputbeforemeas}
\end{eqnarray}

When $k$ photons are detected, the heralded output state is
\begin{eqnarray}
|\Psi_{k}\rangle &=& \sum_{n=0}^{\infty }\sum_{m=0}^{\infty} \gamma^1_{n,\;\ell_1 k_1}\gamma^2_{m, \;\ell_2 k_2}  \sum_{j=0}^{2n+\ell_1-k_1}\gamma_{n, m, j, k-j}\nonumber\\
&&|2n+\ell_1-k_1+2m+\ell_2 -k_2-k\rangle.
 \label{eq:jun1_outputbeforemeas}
\end{eqnarray}
Therefore, odd (even) cats could be generated when $l_{1}-k_{1}+l_{2}-k_{2}-k$ is odd (even). In what follows, we will focus on the simplest case, $l_{1}=l_{2}=0$, $k_{1}=k_{2}=1$, and $k=1$ by considering the experimental feasibility of the scheme.
\subsection{Amplification of Schr\"{o}dinger kitten states}
\begin{figure*} [htbp]
\centering
\vspace{-0.2cm}
\subfigure[]{
\label{fig:subfig:a xi0.68beta1.991} 
\includegraphics[height=2.39in,width=2.95in]{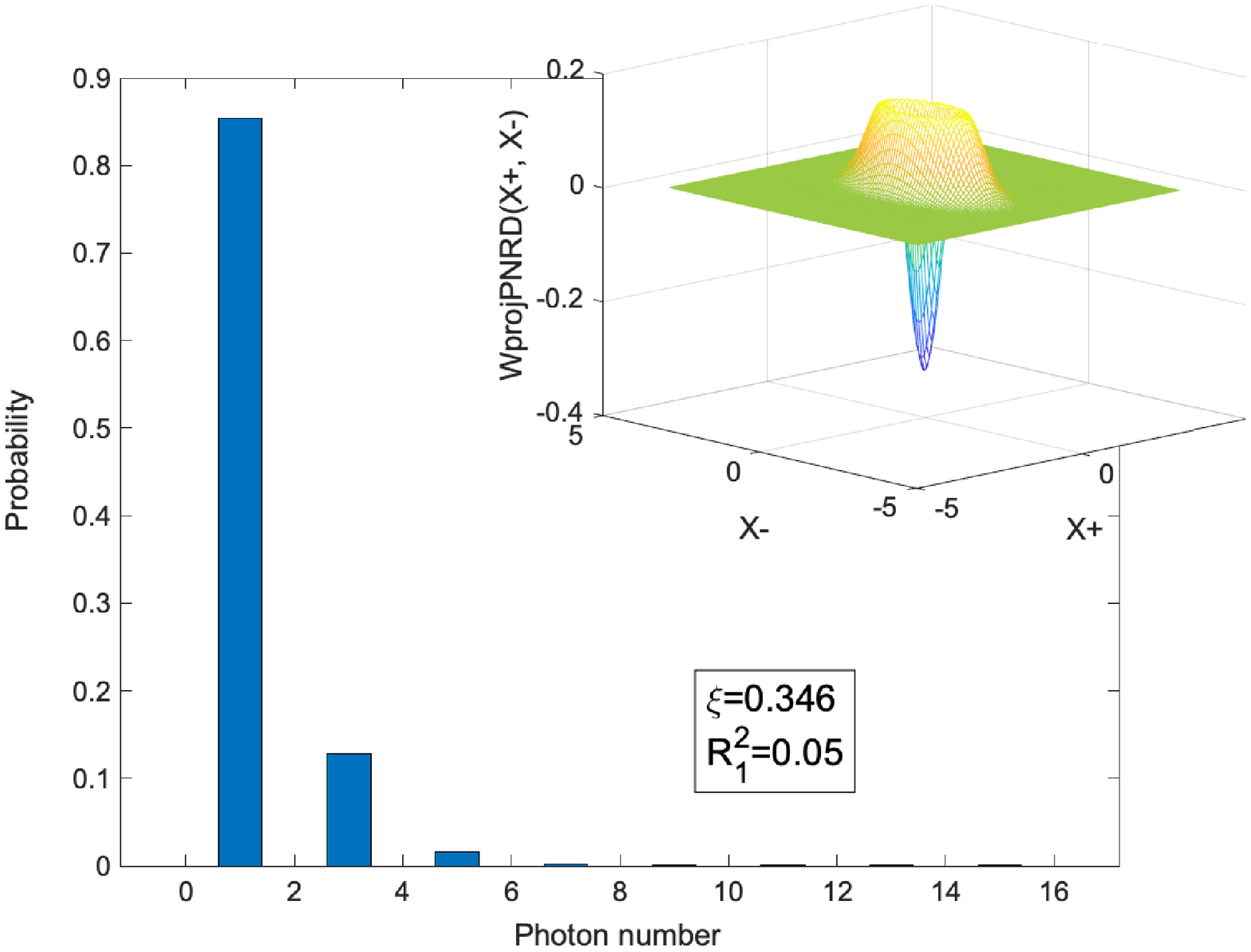}}
\subfigure[]{ \label{fig:subfig:b xi0.7beta2} 
\includegraphics[height=2.39in,width=2.95in]{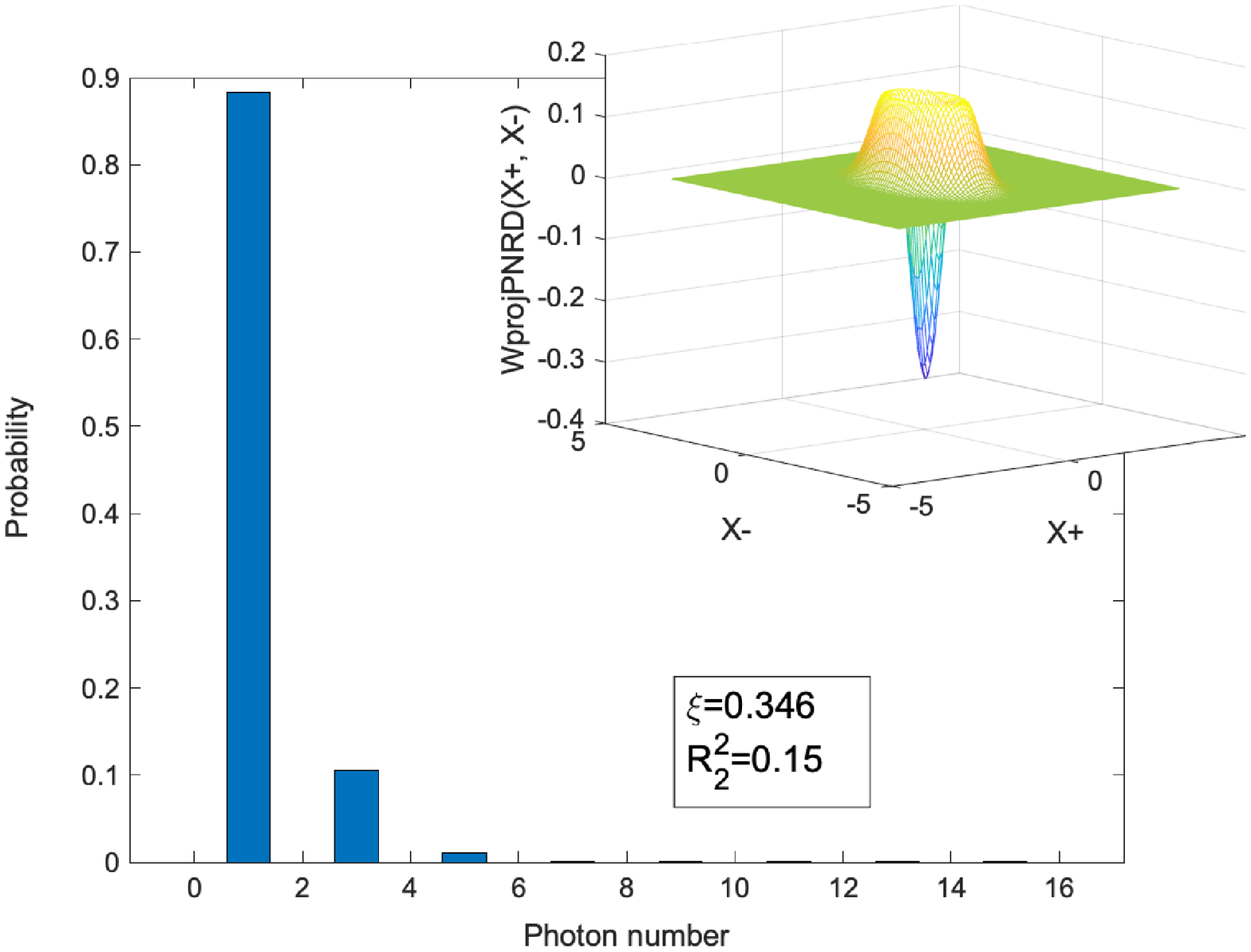}}
\subfigure[]{ \label{fig:subfig:b xi0.7beta2} 
\includegraphics[height=2.39in,width=2.95in]{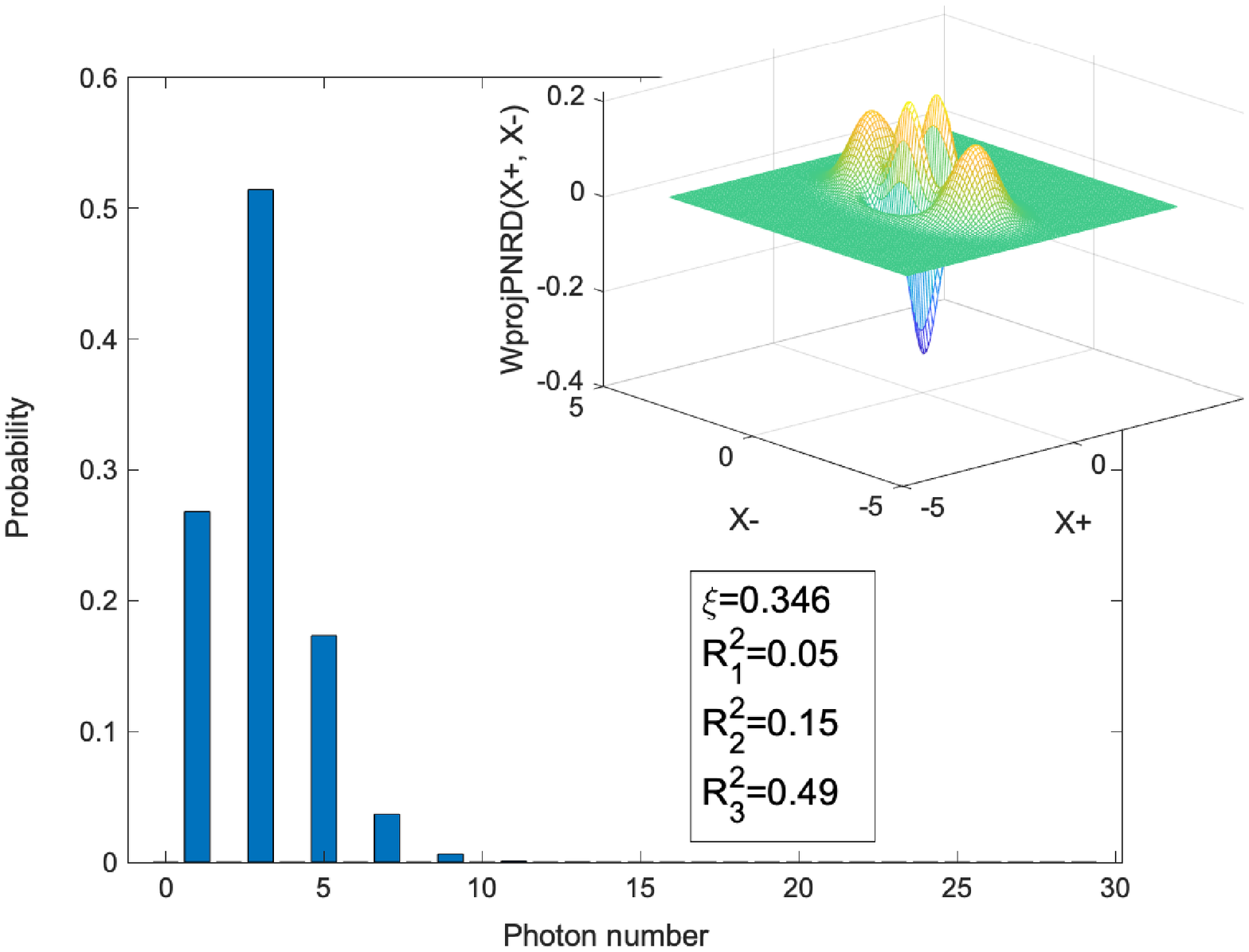}}
\vspace{-0.2cm}
\subfigure[]{
\label{fig:subfig:a xi0.68beta1.991} 
\includegraphics[height=2.39in,width=2.95in]{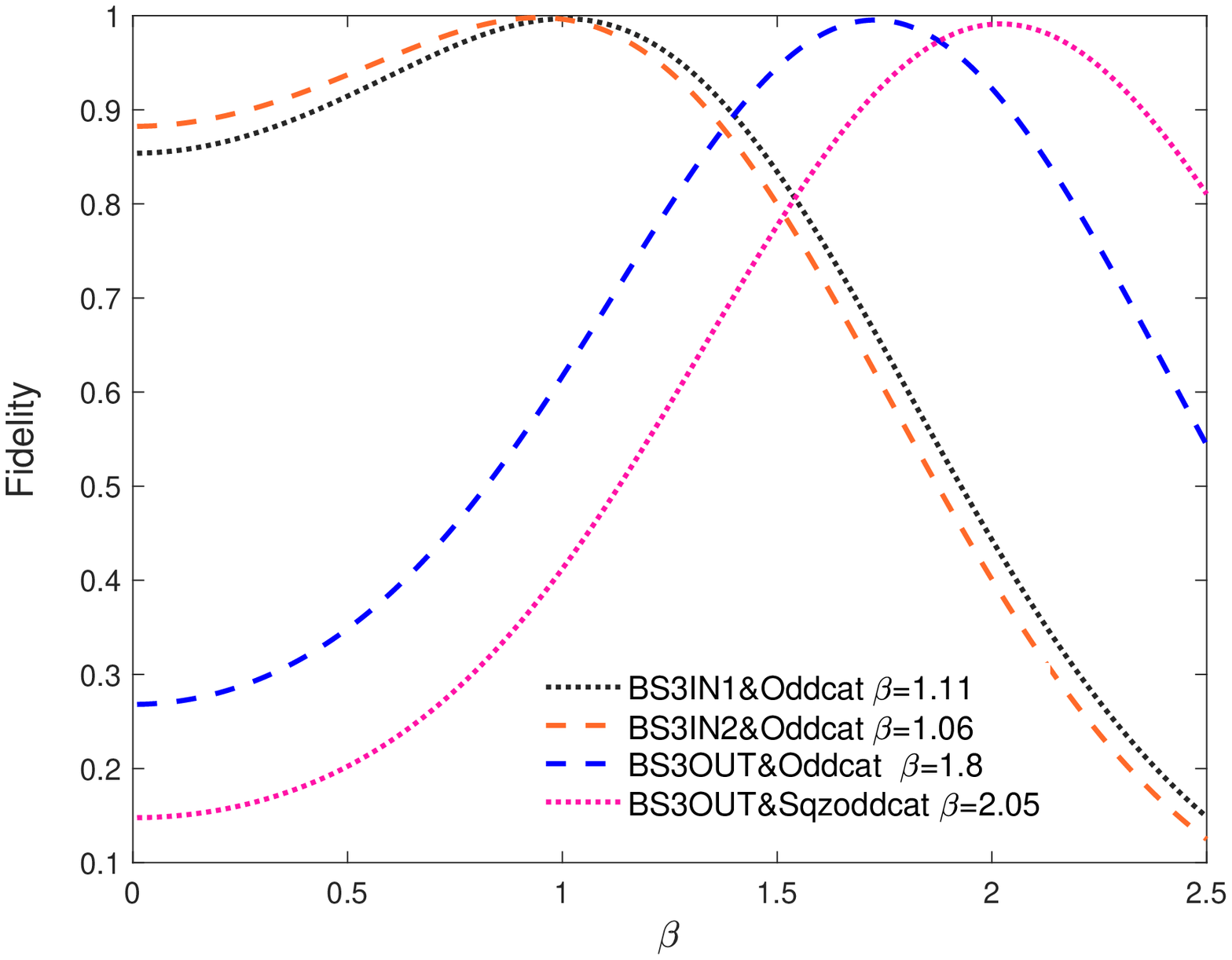}}
\caption{Photon number distribution and Wigner function (insets) of two input kitten states (a) $\beta=1.11$ (b) $\beta=1.06$ \\
 as well as the amplified cat state (c) $\beta=1.80$\\
 (d) Comparison of fidelity variation for input states and output states \\
and the corresponding $\beta$ when $F=99\%$\\
 dotted black line: fidelity between BS3IN1 and an ideal odd cat state \\
 dashed orange line: fidelity between BS3IN2 and an ideal odd cat state\\
dashed blue line: fidelity between the amplified cat state BS3OUT and an ideal odd cat state \\
dotted magenta line: fidelity between the amplified cat state BS3OUT and \\
an ideal cat state squeezed by 1.39 dB}
\label{PND_F_beta_3dB} 
\end{figure*}

As one-photon subtracted squeezed vacuum state is a standard approach to generating kitten states \cite{ScienceOurjoumtsev2006, PRLNielsen2006, OptExpressWakui2007, NatPhotonicsNamekata2010}, which corresponds to the case of $l=0$, $k=1$ in Fig. \ref{l-added-k-sub}, we investigate the amplification effect of the scheme shown in Fig. \ref{l-added-k-sub-two-BS} when the squeezed vacuum state is $-$3 dB corresponding to the squeezing parameter $\xi=0.346$. Fig. \ref{PND_F_beta_3dB} (a), (b) and (c) depict the photon number distribution and Wigner function of two input kitten states and the amplified cat state. The features of odd-photon-number distribution in two input kitten states and the amplified state reveal that the cat parity is remained during the amplification. The fidelity of the input kitten states, BS3IN1 and BS3IN2, as well as the amplified state (BS3OUT) to an ideal odd cat state are represented by the dotted black line, dashed orange line, and dashed blue line in Fig. \ref{PND_F_beta_3dB} (d). It can be seen that two input Schr\"{o}dinger odd kitten states of $\beta=1.11$ and $\beta=1.06$ with $F=99\%$ are amplified to a Schr\"{o}dinger cat state of $\beta=1.80$ and $F=99\%$ with the same parity. In addition, the dotted magenta line in Fig. \ref{PND_F_beta_3dB} (d) implies the fidelity between the amplified cat state and a squeezed cat state, which indicates that the amplified cat state has a fidelity of $F=99\%$ with a cat state of $\beta=2.05$ squeezed by 1.39 dB. The amplification is also revealed by the Wigner function of the input kitten states and the generated state shown in the insets of Fig. \ref{PND_F_beta_3dB} (a)-(c). Different from the Wigner function of the input kitten states, two positive Gaussian peaks related to the individual coherent state components can be clearly observed in the Wigner function of the amplified cat state shown in the inset of Fig. \ref{PND_F_beta_3dB} (c). The notable distance between two Gaussian peaks implying the large size of the generated cat state, which guarantees the distinguishability of two coherent-state components from the non-classical interference fringes.Therefore, different from homodyne-detection heralding scheme \cite{NatphotonSychev2017}, parity-preserving cat state amplification is realized in the scheme as shown in Fig. \ref{l-added-k-sub-two-BS}.

\subsection{Large-size Schr\"{o}dinger cat state generation}
\begin{figure} [htbp]
\centering
\includegraphics[height=2.4in,width=2.95in]{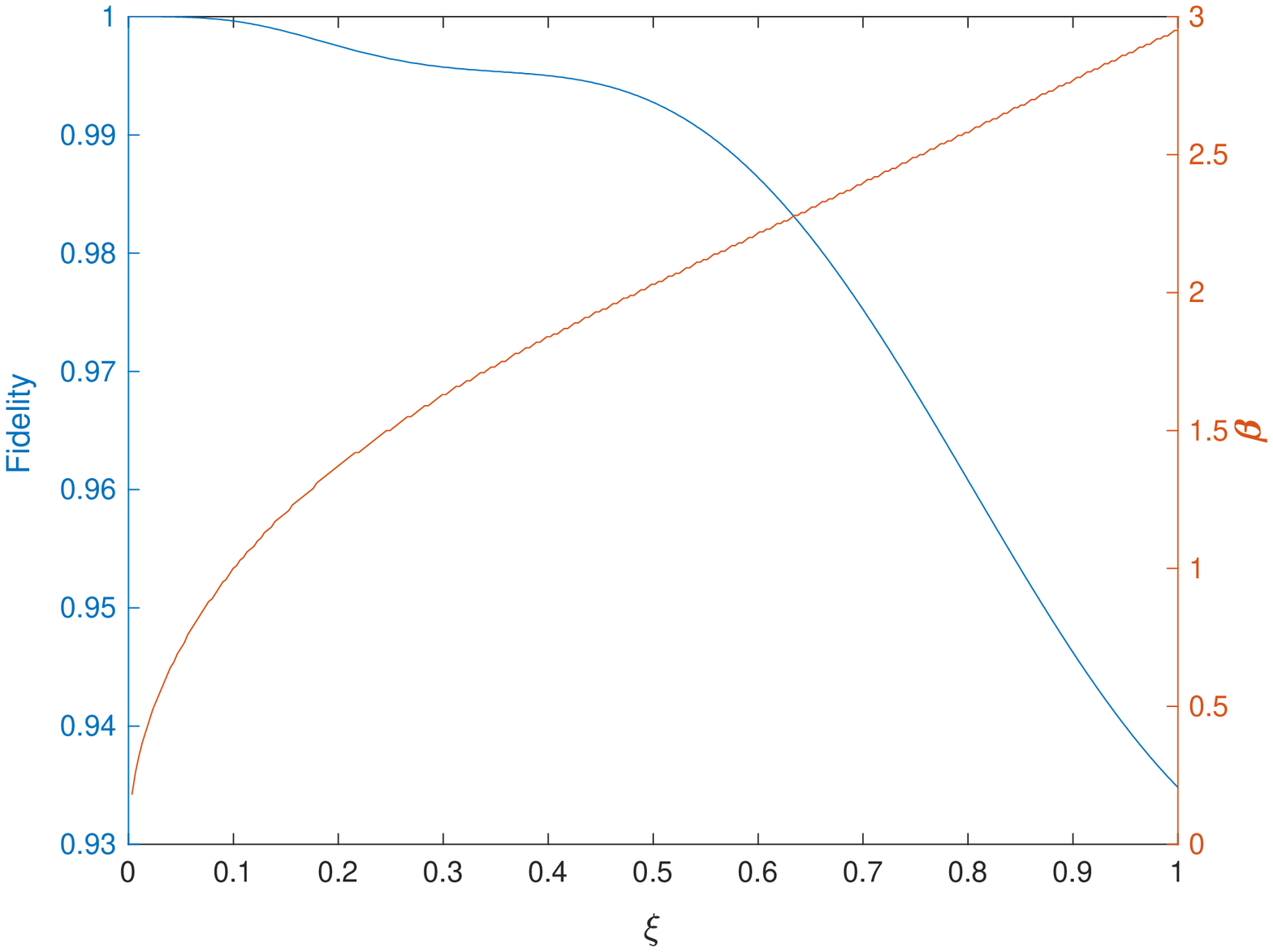}
\caption{Impact of squeezing parameter $\xi$ of the input\\ 
 squeezed vacuum states  on the fidelity $F$ and the size\\
of the amplified cat state $\beta$}
\label{xi_f_beta} 
\end{figure}

\begin{figure*} [htbp]
\centering
\subfigure[]{
\label{fig:subfig:a xi0.68beta1.991} 
\includegraphics[height=2.4in,width=2.95in]{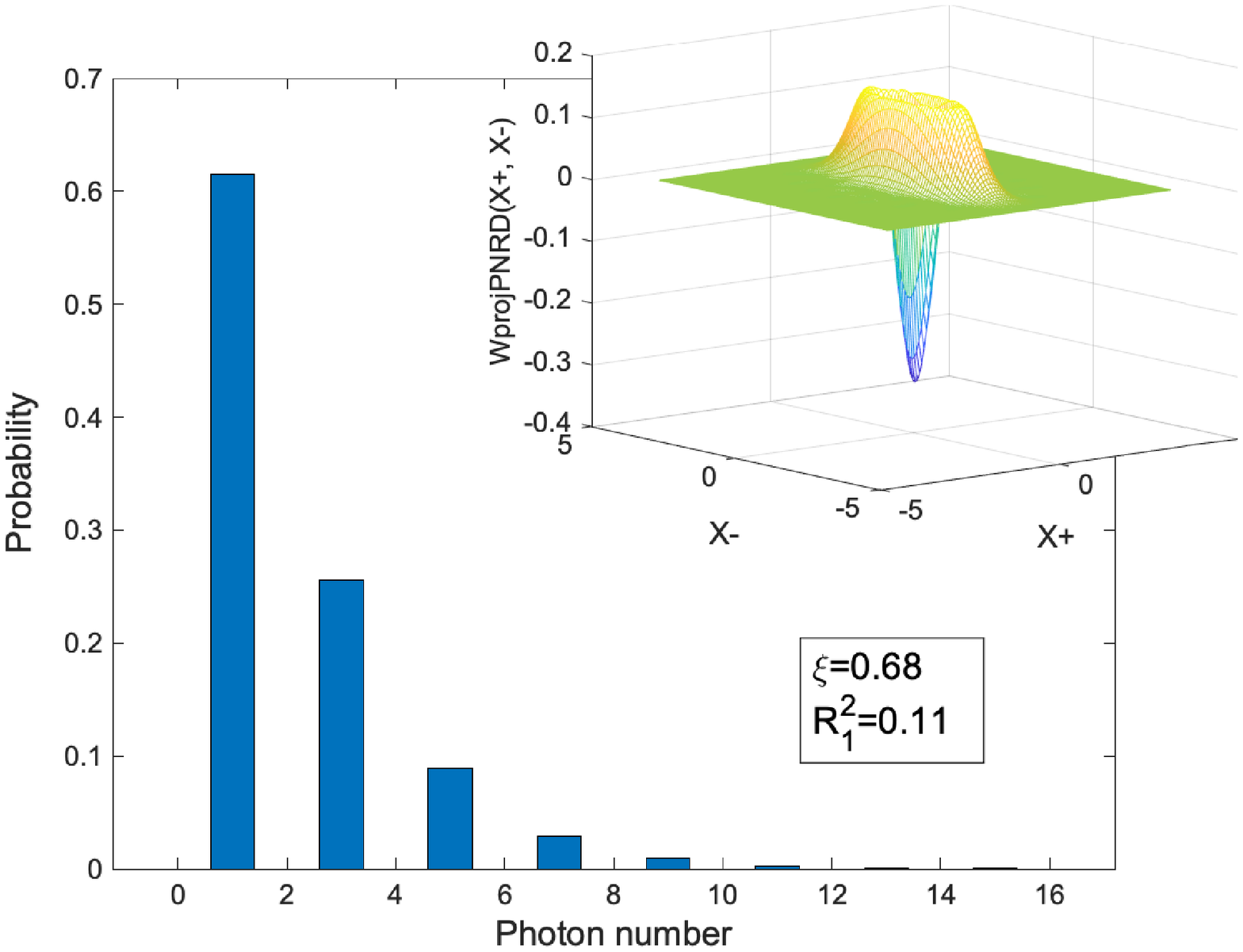}}
\vspace{-2mm}
\subfigure[]{
\label{fig:subfig:c xi0.8} 
\includegraphics[height=2.4in,width=2.95in]{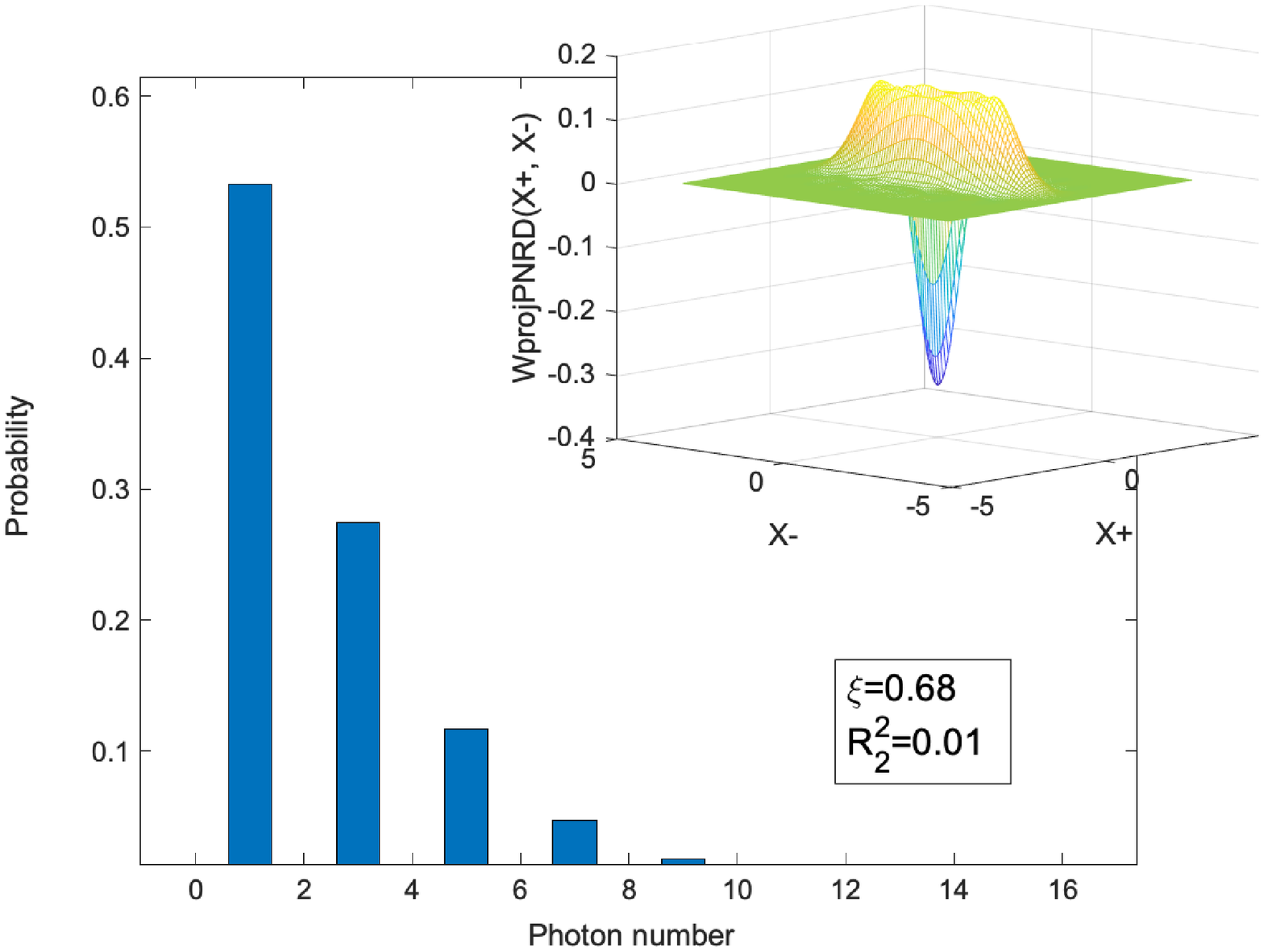}}
\subfigure[]{
\label{fig:subfig:a xi0.68beta1.991} 
\includegraphics[height=2.4in,width=2.95in]{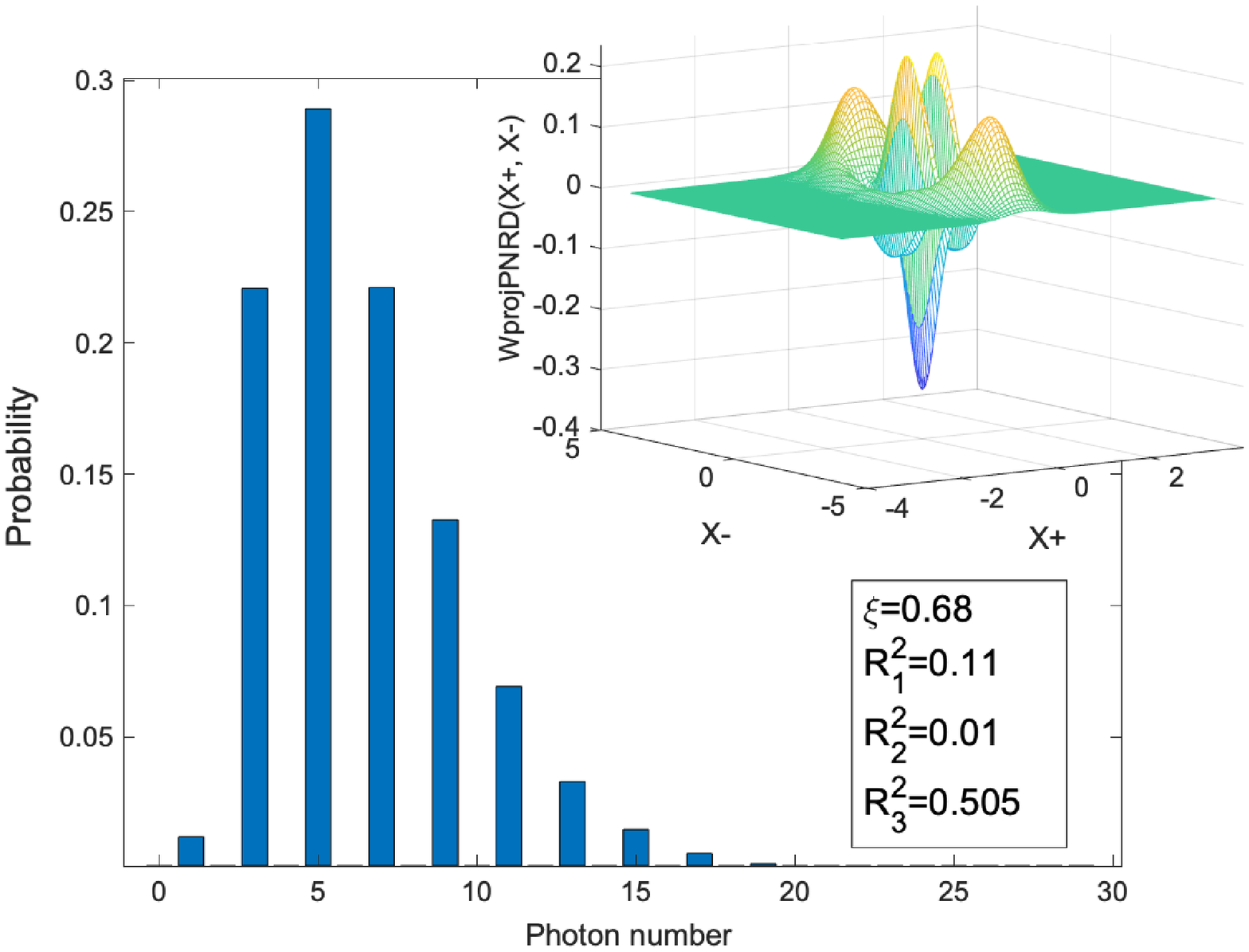}}
\subfigure[]{ \label{fig:subfig:b xi0.7beta2} 
\includegraphics[height=2.4in,width=2.95in]{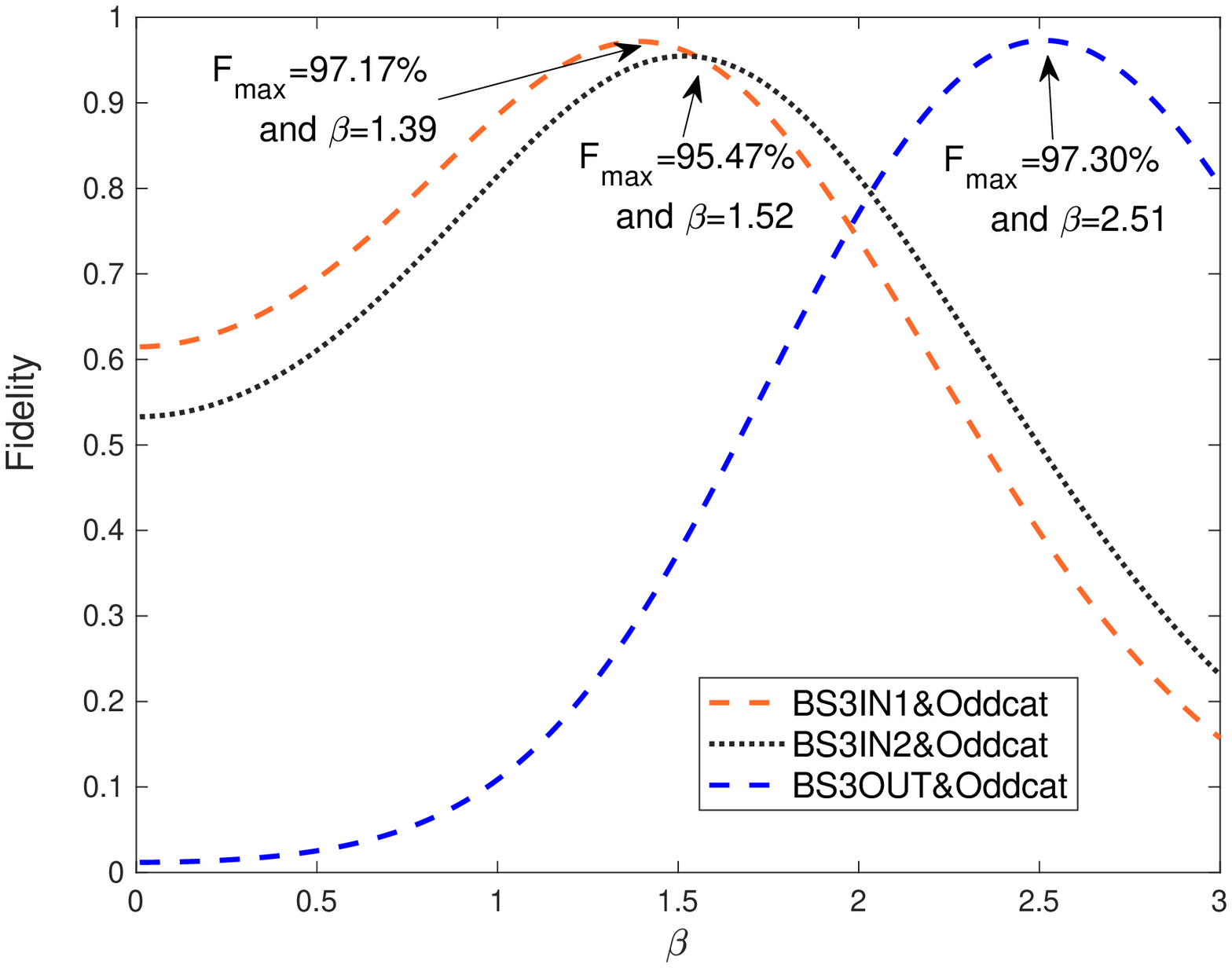}}
\caption{Photon number distribution and Wigner function (insets) of two input kitten states with\\
 (a) $\beta=1.39$, (b) $\beta=1.52$, and the amplified cat state with (c) $\beta=2.51$ \\
(d)Corresponding curves of fidelity varying with $\beta$\\
 dashed orange line: fidelity between BS3IN1 and an ideal odd cat state\\
dotted black line: fidelity between BS3IN2 and an ideal odd cat state\\
dotted blue line: fidelity between the amplified cat state BS3OUT and an ideal odd cat state}
\label{PND_F_beta_xi_68} 
\end{figure*}

The larger the amplitude of a cat state is, the bigger average photon numbers are generated since $\langle n \rangle = |\beta|^2$. Thus squeezed vacuum states with stronger squeezing level are required to generate cat state with $\beta \geq2$. The variation of fidelity and the corresponding $\beta$ of the amplified cat state with the squeezing parameter $\xi$ of the squeezed vacuum states are shown in Fig. \ref{xi_f_beta} when the reflectivities of three BSs are same to the case shown in Fig. \ref{PND_F_beta_3dB}, i.e. $R^2_{1}=0.05$,  $R^2_{2}=0.15$ and $R^2_{3}=0.49$, which indicates that large-size cat states are available with stronger squeezed vacuum states, but the fidelity will decrease. Currently, squeezed vacuum states with the squeezing level as strong as -15 dB has been generated for metrology \cite {PRLVahlbruch2016}. To generate a cat state of $\beta\geq2$ with sufficient fidelity, we select squeezed vacuum states of $-$5.91 dB (corresponding to $\xi=0.68$), which is easy to produce in the lab. Reflectivities of BS1, BS2 and BS3 are numerically optimized with fidelity as $R^2_{1}=0.11$, $R^2_{2}=0.01$ and $R^2_{3}=0.505$, which result in one-photon subtraction from the squeezed vacuum states with success probability of $4.49\%$ and $0.47\%$. The photon number distribution, Wigner function and the corresponding variation of fidelity for two input kitten states and the generated cat state are shown in Fig. \ref{PND_F_beta_xi_68} (a)-(c) and (d). As is shown in Fig. \ref{PND_F_beta_xi_68} (d), when two kitten states of $\beta=1.39$ with $F=97.17\%$ and $\beta=1.52$ with $F=95.47\%$ are input, a large-size Schr\"{o}dinger cat state of $\beta=2.51$ can be generated with the fidelity of $F=97.30\%$. In this case, the success probability is $0.27\%$.  The large size of the generated cat state  is also revealed by the prominent distance between two distinguishable Gaussian peaks in the Wigner function shown in the inset of Fig. \ref{PND_F_beta_xi_68} (c). Therefore, the proposed scheme in Fig. \ref{l-added-k-sub-two-BS} provides an effective approach to generating large-size Schr\"{o}dinger cat states with $\beta\geq2$.

\section{Model with system imperfections} \label{model with imperfection}
\subsection{Imperfections in the kitten states generation}
When a Schr\"{o}dinger kitten state is generated by subtracting one photon from a squeezed vacuum state, it is necessary to consider various imperfections such as impurity of the input squeezed vacuum state, dark counts, detection efficiency and the non-photon-number-resolving ability  of the single photon detector, and so on. The impacts and physical mechanism of all these imperfections are analyzed in detail in \cite{NJPSong2013}. Moreover, with the development of superconducting nanowire single photon detectors (SNSPD), besides the photon-number-resolving ability, single-photon detectors with detection efficiency as high as 98\% and negligible dark counts are available \cite{OpticaReddy2020}. Therefore, we can use SNSPDs as the PNDs in Fig. \ref{l-added-k-sub-two-BS} and just consider the impurity of the squeezed vacuum states (equivalent to the loss) and the detection efficiency of the photon-number-resolving single photon detectors here. According to the Schr\"{o}dinger kitten state model reported in \cite{NJPSong2013}, the generated kitten states from BS1 and BS2 are mixed states,  described as $\rho_{1}$ and $\rho_{2}$, which are functions of the squeezing level $\xi$, the loss $loss_{sqzvac}$ of the squeezed vacuum states, reflectivities of BS1 and BS2, $R^2_{1}$ and $R^2_{2}$, as well as the detection efficiency of the photon-number-resolving PND, $\eta_{PND}$. A general model involving experimental imperfections for the kitten-state amplification scheme is developed. Two input kitten states with experimental imperfections $\rho_1$  and $\rho_2$, can be written as  
\begin{eqnarray}
\rho_1 = \sum_{m,n=0}^\infty \alpha'_{mn} |m\rangle\langle n |,\;
\rho_2 = \sum_{p,q=0}^\infty \beta'_{pq} |p\rangle\langle q|,
   \label{eq:jun1_rho1_2}
\end{eqnarray}
where 
\begin{equation}
\alpha'_{nm} = {\alpha'_{mn}}^{\ast}
\end{equation}
and
\begin{equation}
 \beta'_{qp} = {\beta'_{pq}}^{\ast}
 \end{equation}
 
  Then the input state to BS3 is described as,
\begin{equation}
\rho_\mathrm{in} = \rho_1 \otimes \rho_2 
= \sum_{m,n=0}^\infty  \sum_{p,q=0}^\infty \alpha'_{mn} \beta'_{pq} |mp\rangle{\langle qn|}.
\end{equation}

The output from BS3 is derived as
\begin{widetext}
\begin{eqnarray}
\rho_{\mathrm{out}}
\nonumber
&=&\sum_{m, p=0}^{\infty}\sum_{n, q=0}^{\infty}\frac{\alpha'_{mn} \beta'_{pq}}{\sqrt{m! n! p! q!}} \sum_{j_1=0}^{m}\sum_{i_1=0}^{p} \sum_{j_2=0}^{n}\sum_{i_2=0}^{q} \binom{m}{j_1}\binom{p}{i_1} \binom{n}{j_2}\binom{q}{i_2}(-1)^{i_1+i_2}\nonumber\\
&&T^{p+q+j_1+j_2-i_1-i_2}_{3}R^{m+n+i_1+i_2-j_1-j_2}_{3} \sqrt{(j_1+i_1)! (j_2+i_2)! (m+p-j_1-i_1)! (n+q-j_2-i_2)!}
\nonumber\\
&&|j_1+i_1\rangle |m+p-j_1-i_1\rangle \langle n+q-j_2-i_2|\langle j_2+i_2|.
\end{eqnarray}
\end{widetext}
If we conduct conditional measurement by a Fock state $|k\rangle$ , then the {\it unnormalized} heralded state is
\begin{widetext}
\begin{eqnarray}
\rho_{\mathrm{out},\;k}&&\triangleq\langle k|\rho_\mathrm{out}|k\rangle
\nonumber\\
&&=\sum_{m=0}^\infty \sum_{p = \max(0,k-m)}^\infty \sum_{n=0}^\infty \sum_{q=\max(0,k-n)}^\infty\frac{\alpha'_{mn} \beta'_{pq}}{\sqrt{m! n! p! q!}}\sqrt{(k!)^2 (m+p-k)! (n+q-k)!} \nonumber\\
 &&\sum_{j_1=\max(0, k-p)}^{\min(m,k)} \sum_{j_2=\max(0, k-q)}^{\min(n,k)} \binom{m}{j_1}\binom{p}{k-j_1} \binom{n}{j_2}\binom{q}{k-j_2}  (-1)^{j_1+j_2} T^{p+q+2(j_1+j_2-k)}_{3} 
\nonumber
\\
&&R^{m+n-2(j_1+j_2-k)}_{3} |m+p-k\rangle\langle n+q-k|.   
\label{eq:rhooutk}
\end{eqnarray}
\end{widetext}
\subsection{Imperfections of PND at BS3}

Assume that $m$ photons are detected by the photon-number detector with inefficient detection, as analyzed in \cite{NJPSong2013}, the density matrix of the output state is 
\begin{eqnarray}
\rho_\mathrm{IMPND}(m) = \sum_{k=1}^\infty Q(k|m)\rho_\mathrm{out,\textit{k}}\;,
\end{eqnarray}
where $Q(k|m)$ is the probability that $m$ photons are detected while $k$ photons are actually subtracted and $\rho_\mathrm{out, \;\textit k}$ is the density matrix when $k$ photons are actually subtracted as described in Eq. \eqref{eq:rhooutk}. $Q(k|m)$ can be written as  
\begin{eqnarray}
Q(k|m) = \frac{P(m|k)S(k)}{\sum_i P(m|i)S(i)},
\label{eq:Q(k|m)}
\end{eqnarray}
where $S(k)$ is the probability that $k$ photons have been subtracted \cite{JMO1996} and $P(m|k)$ is the probability that $m$ photons have been detected when $k$ photon are actually subtracted, which are expressed as 
\begin{eqnarray}
S(k)&=&\langle k|\rho_{\mathrm{out}}|k\rangle =\mathrm{Tr}[{\rho_{\mathrm{out},k}}],
 \label{eq:jun1_successprob_nlk}
\end{eqnarray}
which is normalized as $\sum_{k=0}^{\infty}S(k)=1$ and $\rho_{out, k}$ is the unnormalized density matrix,\\ 
\begin{equation}
P(m|k)=\frac{k!\eta_{PNRD}^{m}(1-\eta_{PND})^{k-m}}{m!(k-m)!},\label{equ15}
 \end{equation}
 in which $\eta_{PND}$ is the quantum efficiency of the photon-number-resolving detector \cite{JMO2004}.
 
\subsection{Simulation results}
 \begin{figure*} [htbp]
\centering
\vspace{-10mm}
\subfigure[]{
\label{fig:subfig:a xi0.68beta1.991} 
\includegraphics[height=2.4in,width=2.95in]{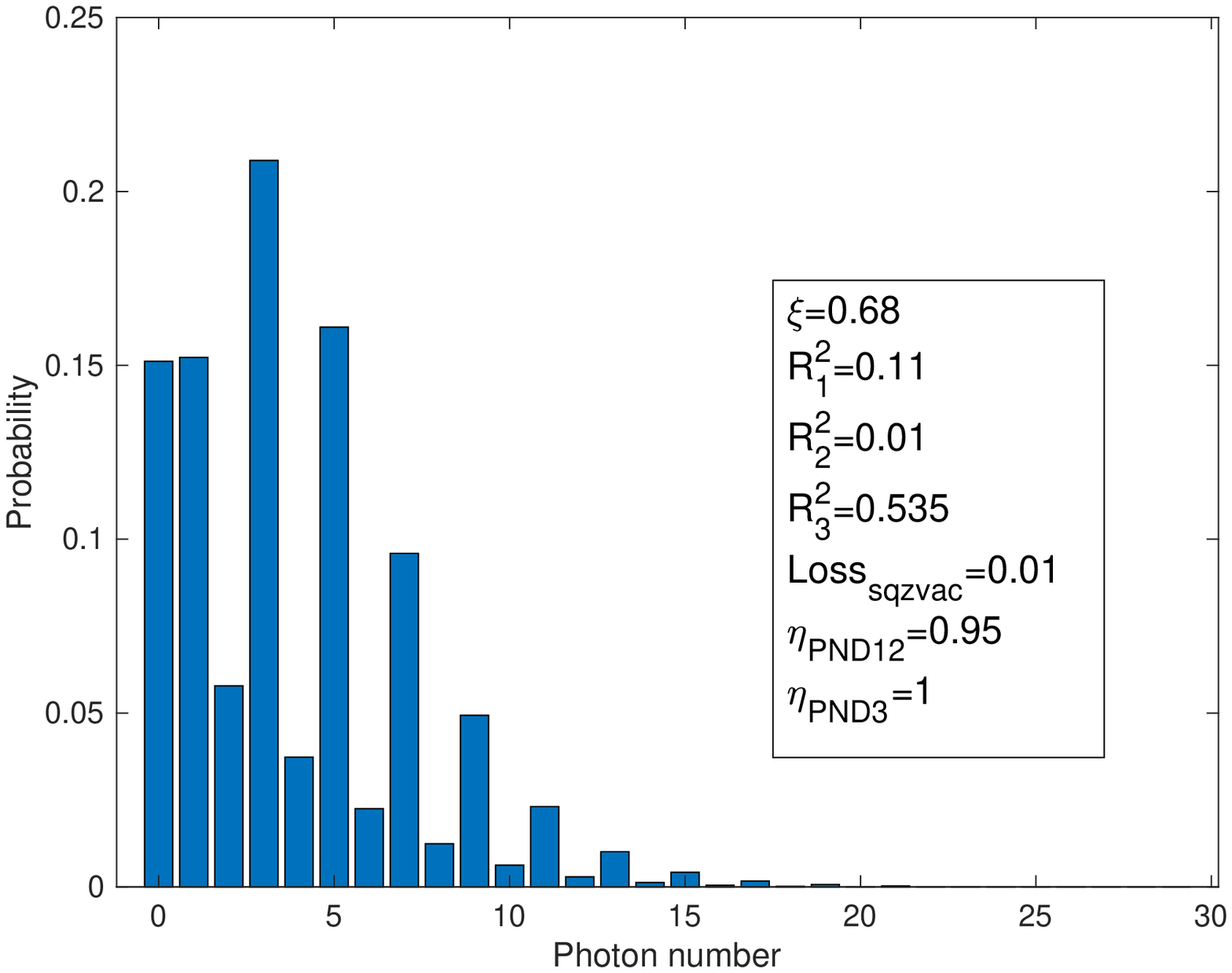}}
\subfigure[]{ \label{fig:subfig:b xi0.7beta2} 
\includegraphics[height=2.4in,width=2.95in]{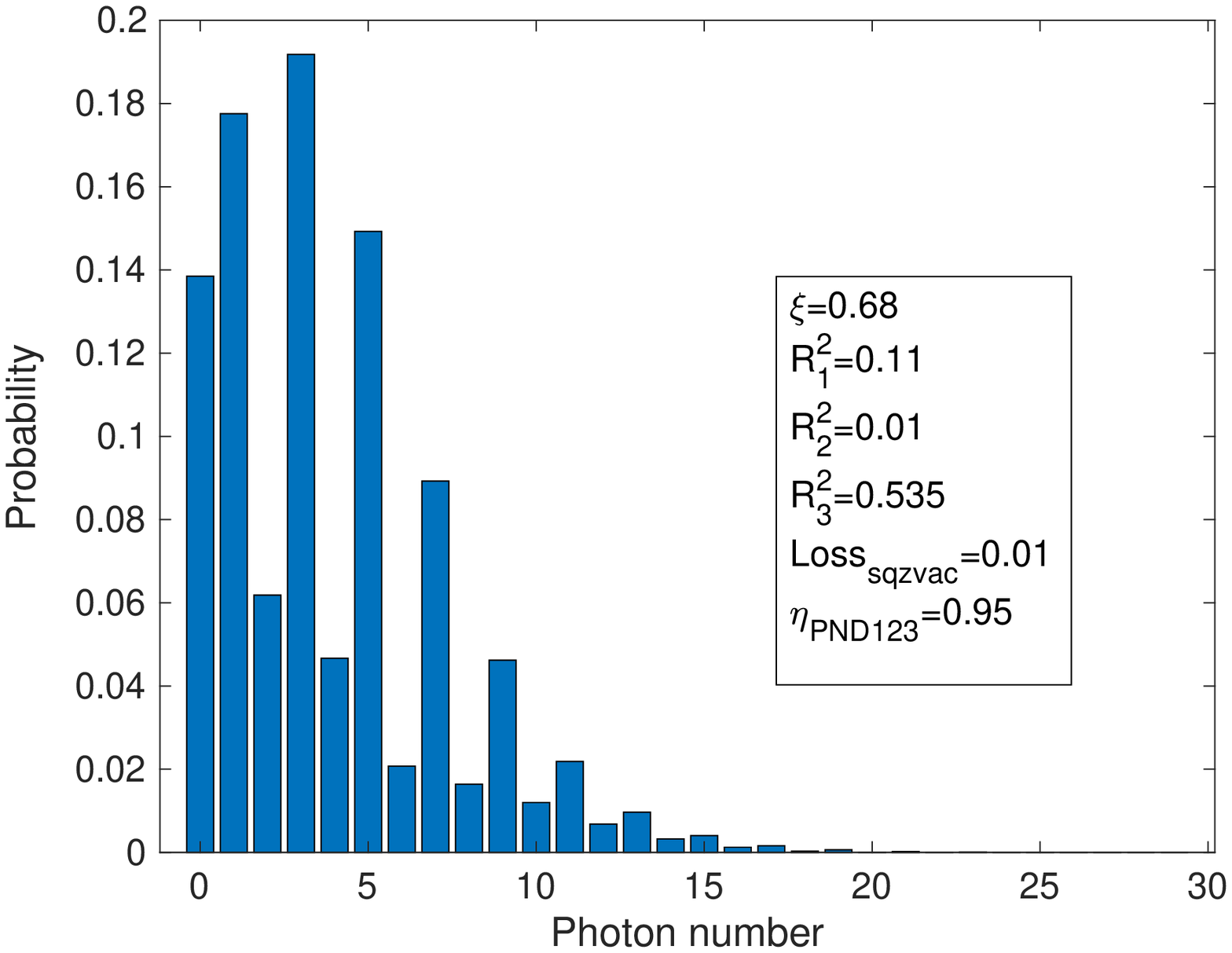}}
\vspace{-3mm}
\subfigure[]{
\label{fig:subfig:a xi0.68beta1.991} 
\includegraphics[height=2.4in,width=2.95in]{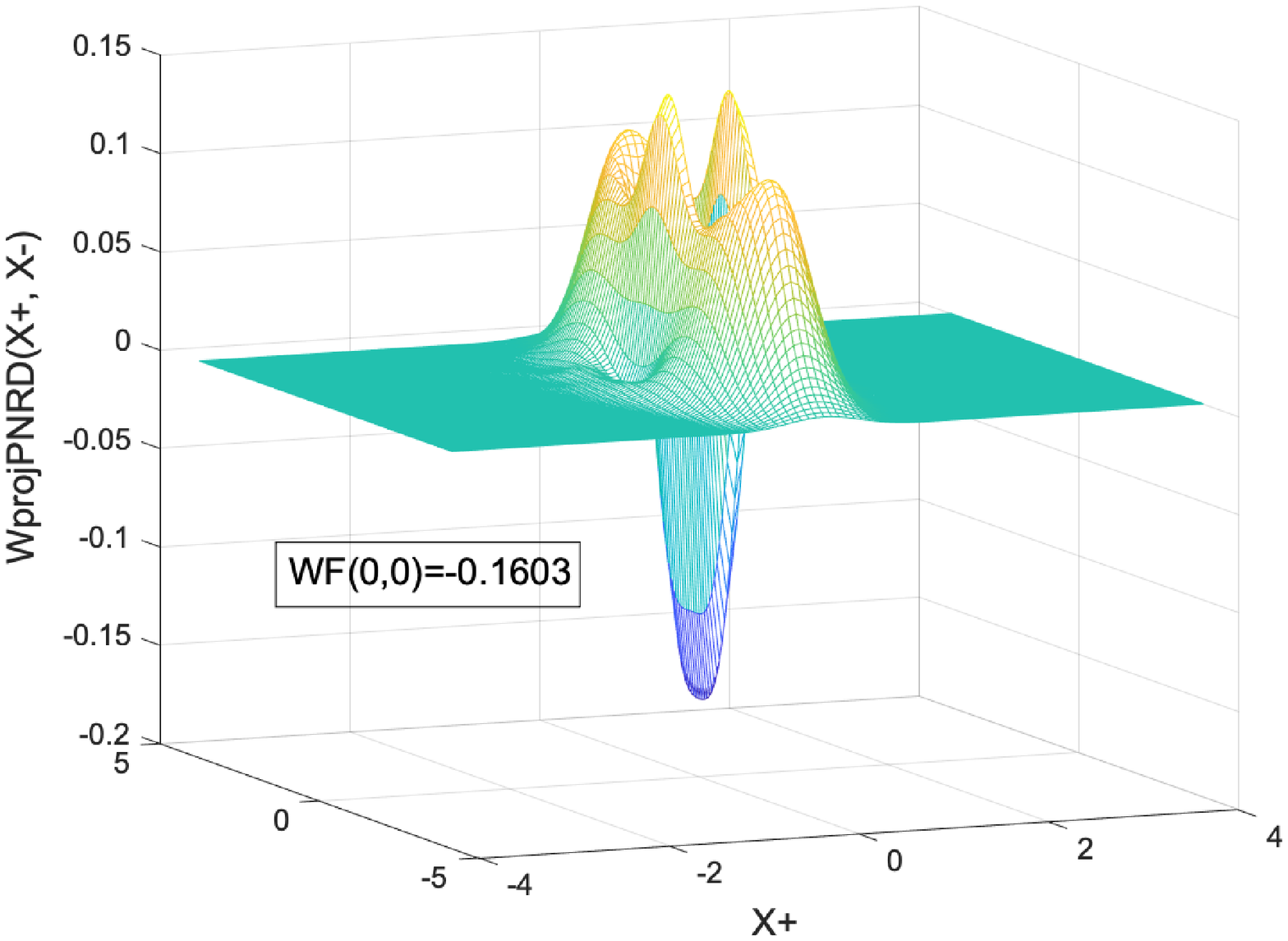}}
\subfigure[]{ \label{fig:subfig:b xi0.7beta2} 
\includegraphics[height=2.4in,width=2.95in]{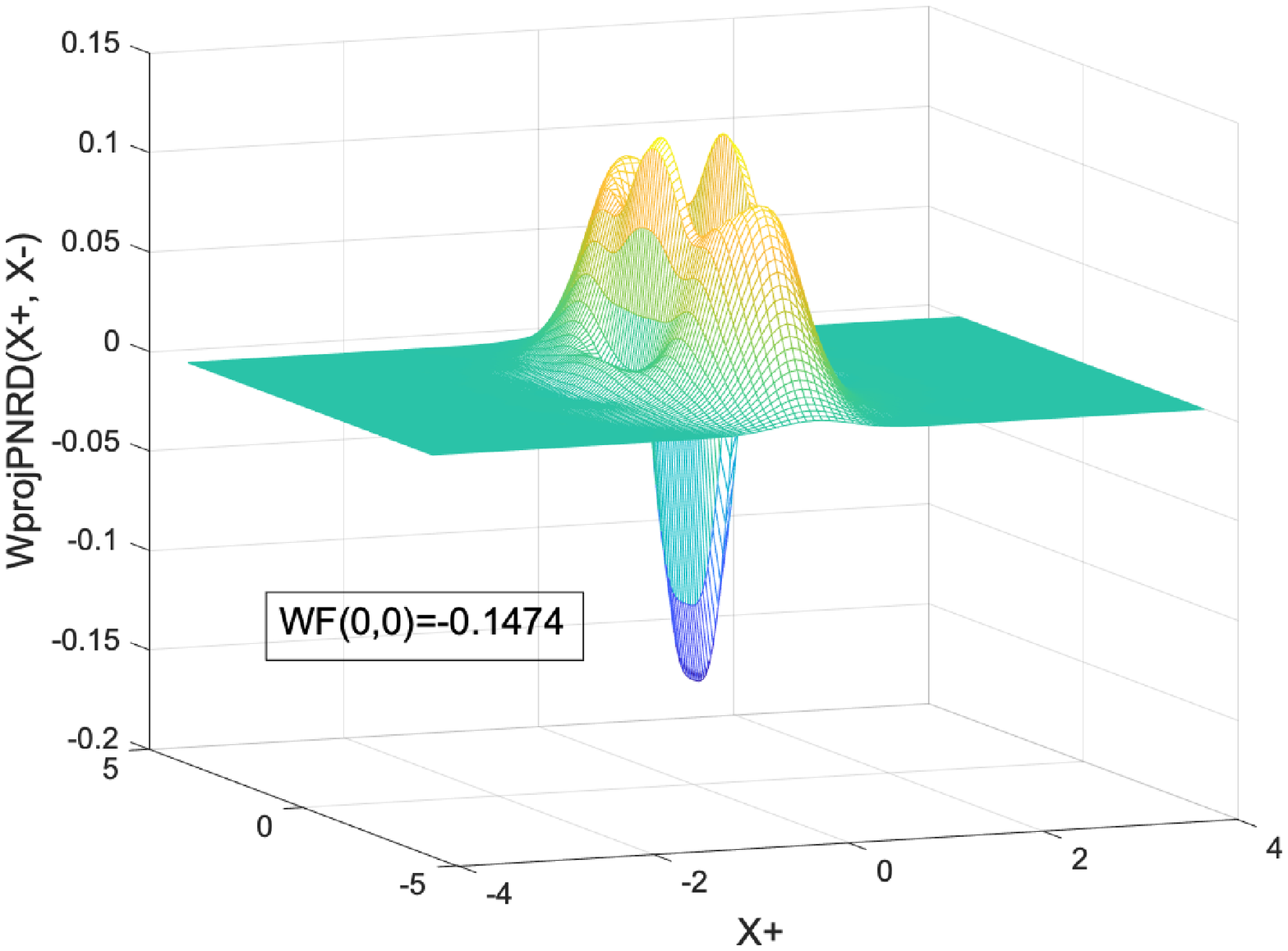}}
\subfigure[]{
\label{fig:subfig:a xi0.68beta1.991} 
\includegraphics[height=2.4in,width=2.95in]{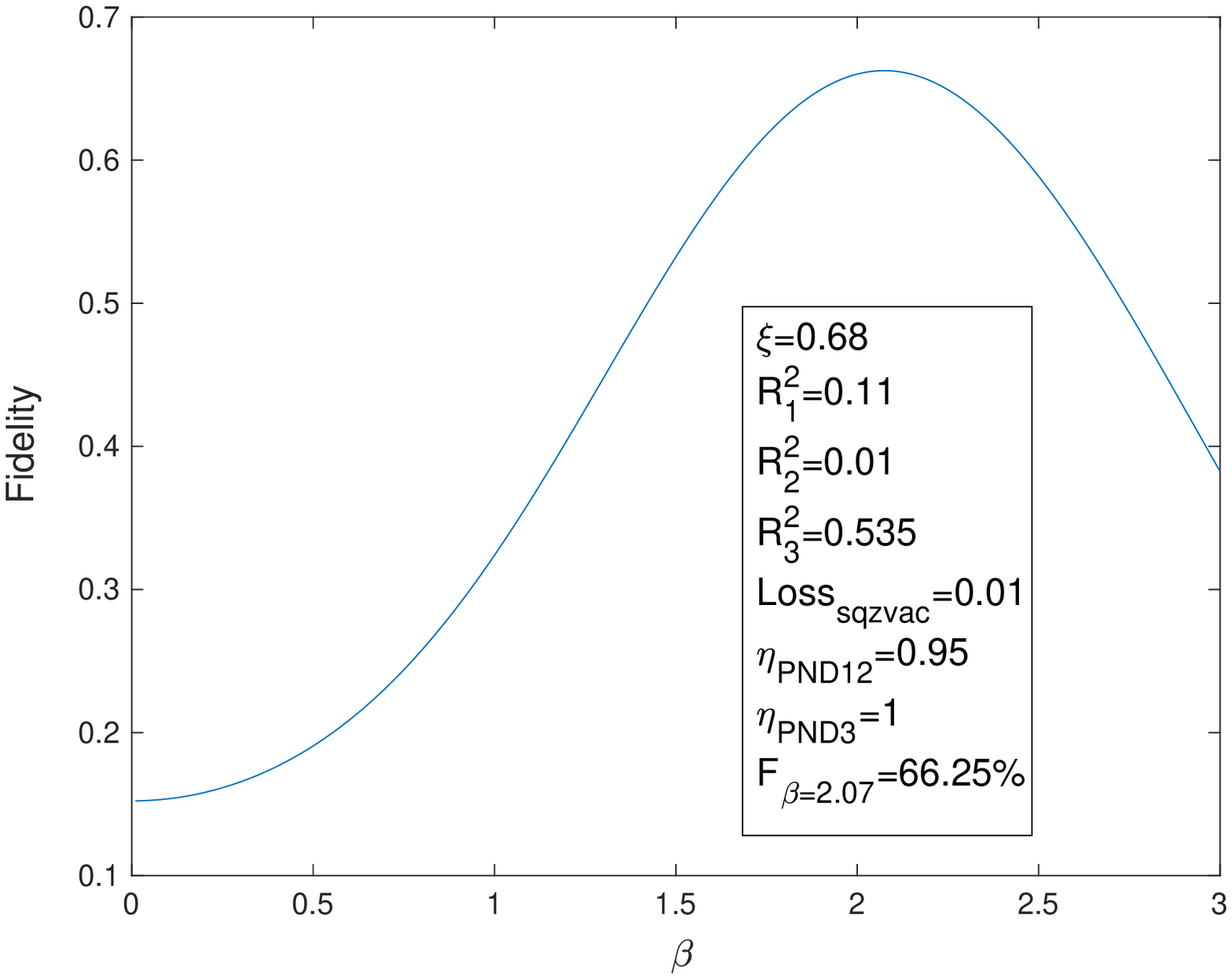}}
\subfigure[]{ \label{fig:subfig:b xi0.7beta2} 
\includegraphics[height=2.4in,width=2.95in]{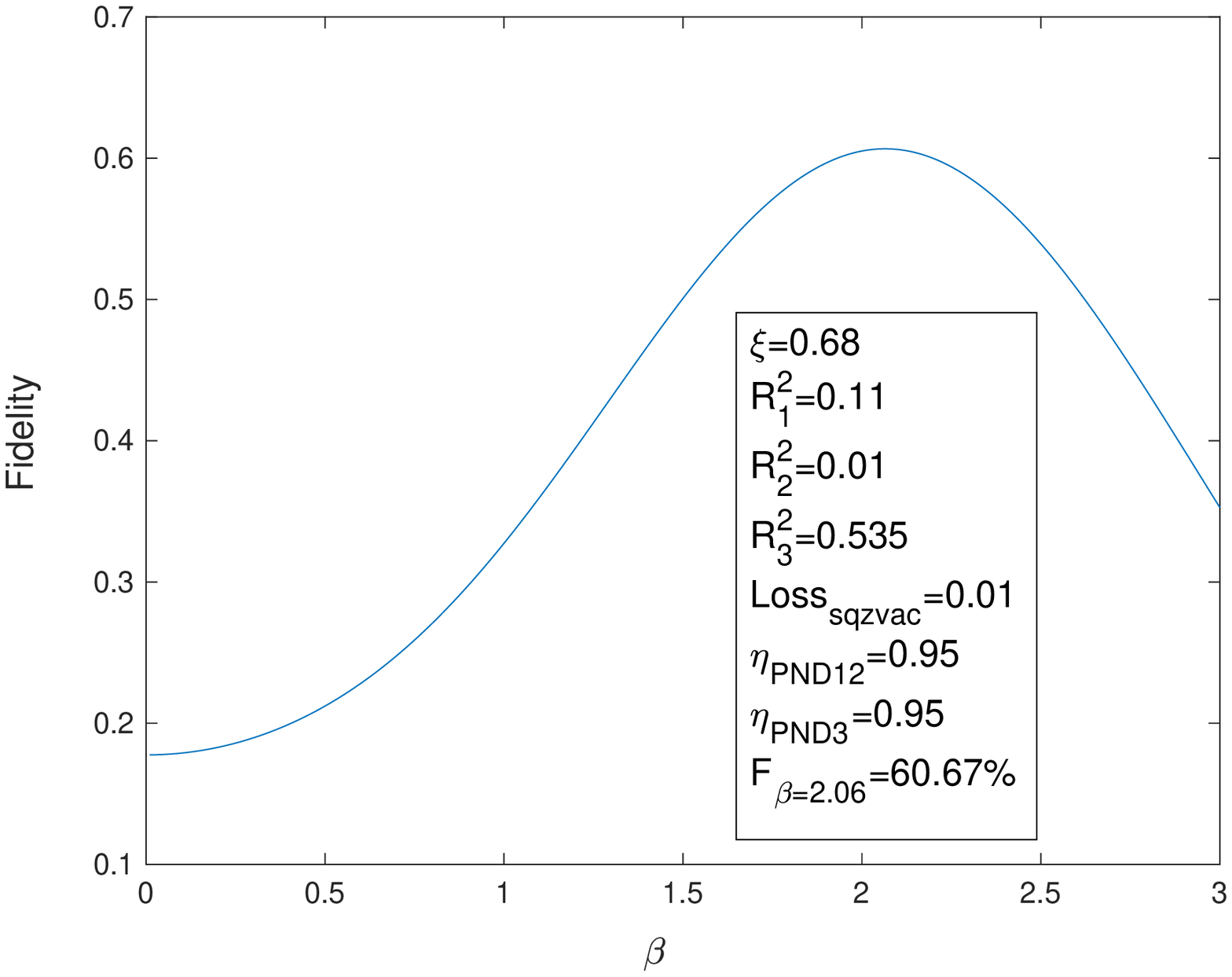}}
 \caption{Amplified cat states heralded with a perfect (left) and imperfect (right) PND3.\\
 (a) and (b) Photon number distribution;  (c) and (d) Wigner function\\
   (e) and (f) Fidelity variation with the amplitude of an ideal cat state}
 \label{PND_losssqz_and_APD_Eff} 
\end{figure*}
As highly-efficient SNSPDs with photon-number-resolving ability and negligible dark counts are commercially available \cite{OpticaReddy2020}, SNSPDs can be used as the PNDs in the cat state amplification scheme proposed. Without losing the generality, we take $\eta_{PND}=0.95$ and $loss_{sqzvac}=0.01$. The photon number distribution and Wigner functions (WF) of the amplified cat state heralded with a perfect and imperfect PND3 are analyzed in the case of  $\xi=0.68$ (i.e. -5.91 dB). Comparing with the photon number distribution of the amplified cat state heralded with an ideal PND3 shown in Fig. \ref{PND_losssqz_and_APD_Eff} (a), P($n$) with $n$=2, 4, 6... in the photon number distribution of the amplified cat state heralded with an imperfect PND3 are increased as shown in Fig. \ref{PND_losssqz_and_APD_Eff} (b). However, significant negativities are kept in both cases as shown in Fig. \ref{PND_losssqz_and_APD_Eff} (c) and (d), which indicate that the odd parity is kept during the amplification in the proposed scheme. Therefore, different from the parity change in \cite{NatphotonSychev2017}, the input odd kitten states are amplified to an odd cat state in the scheme we proposed. Fig. \ref{PND_losssqz_and_APD_Eff} (d) and (e) imply that the amplified cat state of $\beta=2.07$ and $F=66.25\%$ as well as $\beta$=2.06 and $F=60.67\%$ (the fidelity is comparable with that in \cite{NatphotonSychev2017} when the same fidelity definition in the format of square root ($\sqrt{60.67\%}=77.89\%$) is employed.) can be generated with a perfect and imperfect PND3. The success probability is $1.08\%$. It is noted that the generated cat state is sensitive to the loss of the squeezed vacuum states and the detection efficiency of PND3. With the best SNSPD (i.e. $\eta_{PND}=0.98$), cat states with larger size and higher fidelity can be predicted.

\section{Proposal of experimental realization based on a quantum frequency comb} \label{frequency comb-based protocol}

 \begin{figure*} [htbp]
\includegraphics[width=5 in]{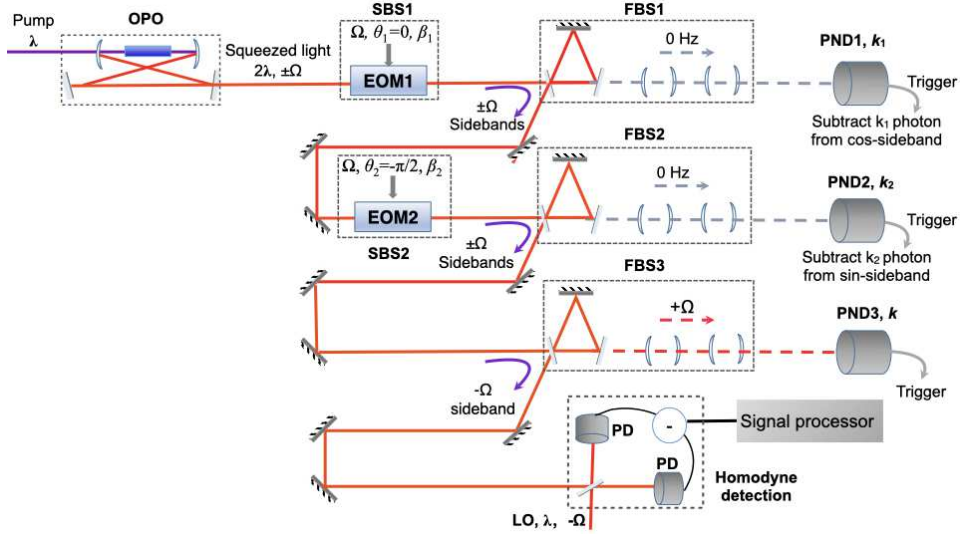}
\caption{Quantum-frequency-comb-based large-size cat state generation scheme \\
 OPO: Optica parametric oscillator; $ \; $SBS: Sideband beam splitter;$\;$  EOM: Electro-optic modulator; \\
 FBS: Frequency beam splitter; $\;$ PND: Photon-number detector; $\;$  LO: Local oscillator;  $\; $PD: Photodetector}
\label{QFC_configuration} 
\end{figure*}

Similar to homodyne-detection heralded cat state amplification scheme \cite{NatphotonSychev2017}, two kitten states are required to obtain a large-size cat state in our proposal. Such kitten states are traditionally generated by subtracting one photon from a squeezed vacuum state  from an optical parametric oscillator (OPO). Thus, two OPOs are required, which makes the system quite complicated. To simplify the experimental setup, we propose a quantum-frequency-comb-based protocol to realize the scheme shown in Fig. \ref{l-added-k-sub-two-BS} by extending the sideband Schr\"{o}dinger cat state generation scheme reported in \cite{PRLSerikawa2018}. As a specific example, we just focus on the case discussed in the previous part, i.e. $l_{1} = l_{2} =0$, $k_{1} = k_{2} = 1$ and $k=1$ in Fig. \ref{l-added-k-sub-two-BS}, which can be realized by the quantum-frequency-comb-based protocol shown in Fig. \ref{QFC_configuration}. It is worth noting that arbitrary-number-photon subtraction can be realized, i.e. $k_1$, $k_2$, and $k$ could be arbitrary integers.   

A series of entangled photon pairs at $\omega_{0}\pm n\Omega$ ($n=1, 2, 3...$),  can be generated by an OPO or OPA, which form a comblike shape \cite{KuesNatPhoton2019, MalteseNPJ2020,ShiOL2020}. For simplicity, we set the baseband at $\omega_{0}=0$ Hz and then the sidebands can be described as $\pm\Omega$. Different from the quantum frequency comb used in quantum teleportation in \cite{SongPRA2014}, the resonant frequency of the OPO in Fig. \ref{QFC_configuration} is set at the sidebands, i.e. $\pm \Omega$, rather than the baseband, which provides a vacuum state at 0 Hz due to the anti-resonance. A double sideband (DBS) mode is defined as $(e^{i\theta}\hat{a}_{+\Omega}+e^{-i\theta}\hat{a}_{-\Omega})/{\sqrt{2}}$, where $\hat{a}_{\pm\Omega}$ is the annihilation operator at frequency $\pm\Omega$ and $\theta$ is an arbitrary phase. A DSB mode can be decomposed into two quadrature-phase components, cos-sideband and sin-sideband,
\begin{eqnarray}
\hat{a}^{\mathrm{cos}}_{\Omega}=\frac{{\hat{a}}_{+\Omega}+{\hat{a}}_{-\Omega}}{\sqrt{2}}\\
\hat{a}^{\mathrm{sin}}_{\Omega}=\frac{{\hat{a}}_{+\Omega}-{\hat{a}}_{-\Omega}}{\sqrt{2}i}
\end{eqnarray}
corresponding to $\theta=0$ and $\theta=-\pi/2$, respectively, from which independent squeezed states are available \cite{ShiOL2020}. Schr\"{o}dinger kitten states can be obtained by subtracting one photon from both cos-sideband and sin-sideband \cite{PRLSerikawa2018}, in which one-photon subtraction from the DBS is realized through a unit composed of an electro-optic modulator (EOM), an optical cavity and a PND.  An EOM plays the role of sideband beam splitter (SBS) via phase modulation. After EOM1, the signal from the OPO, $\hat{a}_{\omega}$, becomes
\begin{equation}
\hat{a}^{\mathrm{mod}}_{\omega}=\sqrt{1-\frac{\beta^2}{2}}\hat{a}_{\omega}+\frac{\beta}{2}(e^{i\theta}{\hat{a}}_{\omega+\Omega}+e^{-i\theta}\hat{a}_{\omega-\Omega}), 
\label {equ:EOM modulation}
\end{equation}
where $\beta\ll1$ indicates the modulation depth and $\theta$ is determined by the modulation phase \cite{PRLSerikawa2018}. While $T^2=1-\frac{\beta^2}{2}$ is equivalent to the transmittance of a beam splitter. An optical cavity following the EOM plays as a frequency beam splitter (FBS) separating and filtering different frequency components by transmitting all baseband and reflecting all sideband signals. Thus the signal from the PND works as a trigger, which heralds \cite{PRLSerikawa2018},
 \begin{equation}
_{trigger}\langle 1|\sim _{sig}\langle 0|(\hat{a}_{0}+\frac{\beta}{\sqrt{2}}\frac{e^{i\theta}{\hat{a}}_{+\Omega}+e^{-i\theta}\hat{a}_{-\Omega}}{\sqrt{2}})
\label{eq:trigger}
\end{equation} 
As a vacuum state at 0 Hz is provided due to the anti-resonance of the OPO, the trigger shown in Eq. (\ref {eq:trigger}) will subtract a photon from the DSB mode with the phase $\theta$. The cos-sideband and sin-sideband can be accessed by selecting the parameters of two EOMs,
 \begin{eqnarray}
&&\theta_{1}=0, \; T^{2}_{1}=1-\frac{\beta_{1}^2}{2}\\
&&\theta_{2}=-\frac{\pi}{2}, \; T^{2}_{2}=1-\frac{\beta_{2}^2}{2},
\end{eqnarray} 
in which $T^2_{i}$ ($i=1$ and 2) of two EOMs in Fig. \ref{QFC_configuration} correspond to the transmittance of BS1 and BS2 shown in Fig. \ref{l-added-k-sub-two-BS}. $T_{i}$  can be adjusted by tunning the modulation depth of EOMs. Therefore, two unbalanced kitten states in both cos-sideband and sin-sideband are generated before entering into FBS3. 
Different from FBS1 and FBS2, FBS3 is designed as a beam splitter of the DBS by separating component $+\Omega$ from -$\Omega$ since we have,
 \begin{eqnarray}
\hat{a}_{+\Omega}=\frac{{\hat{a}}^{\mathrm{cos}}_{\Omega}+i{\hat{a}}^{\mathrm{sin}}_{\Omega}}{\sqrt{2}},\\
\hat{a}_{-\Omega}=\frac{{\hat{a}}^{\mathrm{cos}}_{\Omega}-i{\hat{a}}^{\mathrm{sin}_{\Omega}}}{\sqrt{2}}.
\end{eqnarray} 
A trigger signal at $+\Omega$ is generated from PND3. The function of  BS3 with $T^2_{3} = {\frac{1}{2}}$ in Fig. \ref{l-added-k-sub-two-BS} is realized by shifting the local oscillator in the homodyne detection to the frequency of $-\Omega$. If PND3 is replaced with homodyne detection, the cat amplification scheme based on homodyne heralding reported in \cite{NatphotonSychev2017} could be realized. By designing FBS3 to reflect partial $+\Omega$ and transmit a part of $-\Omega$, $T_{3}$ is adjustable to meet the requirement of cat amplification discussed in Section  \ref{amplificationmodel} and \ref{model with imperfection}. Thus, the implementation of scheme $l_{1}=l_{2}=0$,  $k_{1}=k_{2}=1$, and $k=1$ shown in Fig. \ref{l-added-k-sub-two-BS} is successfully achieved with a quantum frequency comb. 

In addition, if other sidebands, such as $\pm n\Omega$ in the frequency comb are used, multiple large-size cat states are available. It is also potential to amplify the cat state further by an iterative structure. Therefore,  the quantum-frequency-comb-based protocol provides a new approach to producing ``real" coherent-state superposed quantum light source for quantum information processing. \\

\section{Conclusion} \label{conclusion}
In conclusion, a scheme to amplify Schr\"{o}dinger kitten states based on linear operation and conditional measurement with photon number detection is proposed and analyzed. According to the general model of $l$-added-and-$k$-subtracted squeezed vacuum states developed based on tensor operation, the generated kitten states with $l=1$ and $k=2$ as well as $l=3$ and $k=2$ break the limit of $\beta=1.2$ produced in the standard way, i.e. $l=0$ and $k=1$. Combining two kitten states and conducting conditional measurement via photon number detection, Schr\"{o}dinger cat states with $\beta\textgreater2$ are predicted in both theory and practice. A protocol for experimental implementation based on a quantum frequency comb is proposed. Multiple large-size Schr\"{o}dinger cat states are possible to be generated by taking full advantages of the sidebands in a quantum frequency comb, which offers a new way of producing large-scale superposed coherent states to meet the demanding requirements of continuous-variable quantum computation and quantum communication.

\begin{acknowledgments}
This work was supported by National Natural Science Foundation of China under Grant No. 61903316 and Shenzhen Fundamental Research Program under the Grant No. JCYJ20190813165207290.  G. Zhang and X. Wang would like to acknowledge funding from the Hong Kong Research Grant council (RGC) grants (No. 15203619, No. 15208418 and No. 15506619). H. Yonezawa was supported by the Australian Research Council Centre of Excellence for Quantum Computation and Communication Technology (Project No. CE170100012).
\end{acknowledgments}

\bibliography{apssamp}

\begin{thebibliography}{10}
%
%
%
%
%
%
%
%
%
%
%
%
%
%
%
%
%
%
%
%
%
%
%
%
%
%
%
%
%
%
%
%
%
%
%
%
%
%
%

%
%
  \bibitem{PRADakna1997}  M. Dakna, T. Anhut, T.  Opatrny, L. Kn\"{o}ll, and D. G. Welsch, Generation Schr\"{o}dinger-cat-like states by means of conditional measurements on a beam splitter, Phys. Rev. A, {\bf 55},  3184 (1997).
  \bibitem{PRAJeong2003}  H. Jeong, W. Son,  M. S. Kim, D. Ahn, and \ifmmode \check{C}\else \v{C}\fi{}. Brukner, Quantum nonlocality test for continuous-variable states with dichotomic observables, Phys. Rev. A, {\bf 67},  012106 (2003).
   \bibitem{JoptBSangouard2010} N. Sangouard, C. Simon, N. Gisin, J.  Laurat, R.Tualle-Brouri, P. Grangier, Quantum repeaters with entangled coherent states, Soc Am B, {\bf 27}, A137–A145 (2010).   
 \bibitem{PRLBrask2010} J. B. Brask, I. Rigas, E. S. Polzik, U. L. Andersen, A. S. S\o{}rensen, Hybrid long-distance entanglement distribution protocol. Phys. Rev. Lett, {\bf 105}, 160501 (2010).
   \bibitem{NPhotonicsJonas2013} J. S. Neergaard-Nielsen, Y. Eto, C. W. Lee, H. Jeong, and M. Sasaki, Quantum tele-amplification with a continuous variable superposition state, Nat. Photon, {\bf 7},  439-443 (2013).
    \bibitem{PRACochrane1999} P. T. Cochrane, G. J. Milburn, and W. J. Munro, Macroscopically distinct quantum-superposition states as a bosonic code for amplitude damping, Phys. Rev. A, {\bf 59}, 2631 (1999)
      %
    \bibitem{PRARalph2003}  T. C. Ralph, A. Gilchrist, and G. J. Milburn, Quantum computation with optical coherent states, Phys. Rev. A, {\bf 68},  042319 (2003).
    
        \bibitem{OLHastrup2020}  J. Hastrup, J. S. Neergaard-Nielsen, and U. L. Andersen, Deterministic generation of a four-component optical cat state, Opt. Lett, {\bf 45},  640-643 (2020).
    
  \bibitem{PRLLund2007} A. P. Lund, T. C. Ralph, and H. L. Haselgrove, Fault-tolerant linear optical quantum computing with small-amplitude coherent states, Phys. Rev. Lett, {\bf 100},  030503 (2008). 
  
 \bibitem{JOptBGilchristy2004} A. Gilchrist, K. Nemoto, W. Munro, T. Ralph, S. Glancy, S. Braunstein, and G. Milburn, Schr\"{o}dinger cats and their power for quantum information processing, J. Opt. B: Quantum and Semiclassical Optics, {\bf 6},  S828 (2004).
 
  \bibitem{PRAMunro2002}  W. J. Munro, K. Nemoto, G. J. Milburn, and S. L. Braunstein, Weak-force detection with superposed coherent states, Phys. Rev. A, {\bf 66},  023819 (2002).
 \bibitem{PRLMunro2011}  J. Joo, William J. Munro, and T. P. Spiller, Quantum metrology with entangled coherent states, Phys. Rev. Lett, {\bf 107},  083601 (2011).
 
  
 \bibitem{ScienceOurjoumtsev2006}  A. Ourjoumtsev, R. Tualle-Brouri, J. Laurat, P. Grangier, Generating Optical Schr\"{o}dinger Kittens for Quantum Information Processing, Science, {\bf 312},  83-86 (2006).
   \bibitem{PRLNielsen2006} J. S. Neergaard-Nielsen, B. Melholt Nielsen, C. Hettich, K. M\o{}lmer, and E. S. Polzik, Generation of a Superposition of Odd Photon Number States for Quantum Information Networks, Phys. Rev. Lett, {\bf 97}, 083604 (2006).
  \bibitem{NatPhotonicsNamekata2010} N. Namekata, Y. Takahashi, G. Fujii, D. Fukuda, S. Kurimura, and S. Inoue, Non-Gaussian operation based on photon subtraction using a photon-number-resolving detector at a telecommunications wavelength, Nat. Photon, {\bf 4}, 655-660 (2010).

 \bibitem{OptExpressWakui2007}K. Wakui, H. Takahashi, A. Furusawa, and M. Sasaki, Photon subtracted squeezed states generated with periodically poled KTiOPO4, Opt. Express, {\bf 15}, 3568-3574 (2007).

  \bibitem{PRLSerikawa2018}T. Serikawa, J. Yoshikawa, S.Takeda, H. Yonezawa, T. C. Ralph, E. H. Huntington, and A. Furusawa, Generation of a cat state in an optical sideband, Phys. Rev. Lett, {\bf 121}, 143602(2018).
  
  
  
  
  
\bibitem{PRLLund2004} A. P. Lund, H. Jeong,  T. C. Ralph, and M. S. Kim, Conditional production of superpositions of coherent states with inefficient photon detection, Phys. Rev. A, {\bf 70}, 020101(R) (2004).
\bibitem{PRAAgata2008} A. M. Branczyk and T. C. Ralph, Teleportation using squeezed single photons, Phys. Rev. A, {\bf 78},  052304 (2008).


\bibitem{NatureOurjoumtsevl2007} A. Ourjoumtsev, H. Jeong, R. Tualle-Brouri, P. Grangier, Generation of optical Schr\"{o}dinger cats from photon number states, Nature, {\bf 448}, 784-786 (2007).


\bibitem{NatphotonSychev2017}D. V. Sychev, A. E. Ulanov, A.  A. Pushkina, M. W. Richards, I. A. Fedorov, and A. I. Lvovsky, Enlargement of optical Sch\"{o}dinger's cat states, Nat. Photon, {\bf 11}, 379-383 (2017).


\bibitem{PRLTakahashi2008} H. Takahashi, K. Wakui, S. Suzuki, M. Takeoka, K. Hayasaka, A. Furusawa, and M. Sasaki, Generation of Large-Amplitude Coherent-State Superposition
via Ancilla-Assisted Photon Subtraction, Phys. Rev. Lett, {\bf 101}, 233605 (2008).

\bibitem{PRAMasahirohiroki2008} T. Masahiro, T. Hiroki, and S. Masahide, Large-amplitude coherent-state superposition generated by a time-separated two-photon subtraction from a continuous-wave squeezed vacuum, Phys. Rev. A,  {\bf 77}, 062315 (2008).

\bibitem{PRANielsen2007} A. E. B. Nielsen, and K. Molmer, Transforming squeezed light into a large-amplitude coherent-state superposition, Phys. Rev. A, {\bf 76},  043840 (2007).

 \bibitem{PRAGerrits2010}T. Gerrits, S. Glancy,T. S. Clement, B. Calkins, A. E. Lita,  A. J. Miller, A. L. Migdall, S. Woo Nam, R. P. Mirin, and E. Knill, Generation of optical coherent-state superposition by number-resolved photon subtraction from the squeezed vacuum, Phys. Rev. A, {\bf 82},  031802(R) (2010).











\bibitem{PRALaghaout2013} A. Laghaout, J. S. Neergaard-Nielsen, I. Rigas, C. Kragh, A. Tipsmark, and U. L. Andersen, Amplification of realistic Schr\"{o}dinger-cat-state-like states by homodyne heralding, Phys. Rev. A, {\bf 87}, 043826 (2013).





\bibitem{PRATakase2021} K. Takase,  J. Yoshikawa, W. Asavanant, M. Endo, and A. Furusawa, Generation of optical  Schr\"{o}dinger cat states by general photon subtraction,  Phys. Rev. A, {\bf 103}, 013710 (2021).
\bibitem{NJPLEaton2019} M. Eaton, R. Nehra, and O. Pfister, Non-Gaussian and Gottesman-Kitaev-Preskill state preparation by photon catalysis, New. J. Phys, {\bf 21}, 113034, (2019).
\bibitem{ScientificreportMilheev2019} E. V. Mikheev, A. S. Pugin, D.  A. Kuts, S.  A. Podoshvedov, and N. Ba An, Efficient production of large-size optical Schr\"{o}dinger cat states, Sci. Rep,  {\bf 9}, 14301 (2019).



\bibitem{IEEETRANONAUTOCONTROLZhang2013} G. Zhang, and M. James. On the response of quantum linear systems to single photon input fields,IEEE Trans. Automat. Contr,  {\bf 58}, 1221-1235 (2013).

\bibitem{Automaticzhang2014} G. Zhang, Analysis of quantum linear systems{'} response to multi-photon states, Automa,  {\bf 2}, 442-451 (2014).

\bibitem{Automaticzhang2017} G. Zhang, Dynamical analysis of quantum linear systems driven by multi-channel multi-photon states, Automa,  {\bf 83}, 186-198 (2017).



\bibitem{JOMJozsa1994} R. Jozsa,  Fidelity for Mixed Quantum States, J. Mod. Opt, {\bf 41}, 2315-2323 (1994).




\bibitem{PRLVahlbruch2016} H. Vahlbruch, M. Mehmet, K. Danzmann, and R. Schnabel, Detection of 15 dB Squeezed States of Light and Their Appli- cation for the Absolute Calibration of Photoelectric Quantum Efficiency, Phys. Rev. Lett. 117, 110801 (2016).
 



\bibitem{NJPSong2013} H. Song, K. B. Kuntz, and E. H. Huntington, Limitations on the quantum non-Gaussian characteristic of Schr\"{o}dinger kitten state generation, New. J. Phys, {\bf 15}, 023042 (2013).

\bibitem{OpticaReddy2020} D. V. Reddy, R. R. Nerem, S. W. Nam, R. P. Mirin, and V. B. Verma, Superconducting nanowire single-photon detectors
with 98\% system detection efficiency at 1550 nm, Optica, {\bf 7}, 1649-1652 (2020).

 \bibitem{JMO1996} B. Masahi, Photon statistics of conditional output states of lossless beam splitter, J. Mod. Opt. {\bf 43} 1281-1303 (1996).
  \bibitem{JMO2004}H. Lee, U. Yurtsever, P. Kok, G. M. Hockney, C. H. Adami, S. L. Braunsterin and J. Dowling, Towards photostatistics from photon-number discriminating
detectors, J. Mod. Opt. {\bf 51} 1517-1528 (2004).
  
%
%
%


  \bibitem{KuesNatPhoton2019}M. Kues, C. Reimer, J. M. Lukens, W. J. Munro, A. M. Weiner, D. J. Moss, and R. Morandotti, Quantum optical microcombs, Nat. Photon, {\bf 13}, 170-179, (2019).

  \bibitem{MalteseNPJ2020} G. Maltese, M. I. Amanti, F. Appas, G. Sinnl, A. Lemaître, P. Milman, F. Baboux, and S. Ducci, Generation and symmetry control of quantum frequency combs,  NPJ Quantum Inf, {\bf 6}, 13, (2020).   
    \bibitem{ShiOL2020} S. Shi, Y. Wang, L. Tian, J. Wang, Xiao. Sun, and Y. Zheng, Observation of a comb of squeezed states with a strong squeezing factor by a bichromatic local oscillator, Opt. Lett, {\bf 45}, 2419-2422 (2020). 
    \bibitem{SongPRA2014} H. Song, H. Yonezawa, K. B. Kuntz, M. Heurs, and E. H. Huntington, Quantum teleportation in space and frequency using entangled pairs of photons
from a frequency comb, Phys. Rev. A, {\bf 90}, 042337 (2014).
\end{thebibliography}

\end{document}